\input pipi.sty
\input epsf.sty
\input psfig.sty
\magnification1000
\vsize=21cm
\def\otightboxit{\relax}
\raggedbottom

\nopagenumbers
\rightline{April, 2006}
\rightline\timestamp
\rightline{FTUAM 06-1}
\rightline{hep-ph/0603170}

\bigskip
\hrule height .3mm
\vskip.6cm
\centerline{{\bigfib The pion-pion scattering amplitude. II:}} 
\centerline{{\bigfib Improved analysis above $\bar{K}K$
threshold\vphantom{\Bigg|}}}
\medskip
\centerrule{.7cm}
\vskip1.7cm
\setbox1=\vbox{\hsize65mm {\noindent\fib R.~Kami\'nski} 
\vskip .1cm
\noindent{\addressfont Department of Theoretical Physics\hb
Henryk Niewodnicza\'nski Institute\hb of Nuclear Physics,\hb
Polish Academy of Sciences,\hb
31-242, 
Krak\'ow, Poland,}}
\medskip
\setbox8=\vbox{\hsize65mm {\noindent\fib J. R. Pel\'aez} 
\vskip .1cm
\noindent{\addressfont Departamento de F\'{\i}sica Te\'orica,~II\hb
 (M\'etodos Matem\'aticos),\hb
Facultad de Ciencias F\'{\i}sicas,\hb
Universidad Complutense de Madrid,\hb
E-28040, Madrid, Spain}}
\line{\hfil\box1\hfil\box8\hfil}
\smallskip
\setbox7=\vbox{\hsize65mm \fib and} 
\centerline{\box7}
\smallskip
\setbox9=\vbox{\hsize65mm {\noindent\fib F. J. 
Yndur\'ain} 
\vskip .1cm
\noindent{\addressfont Departamento de F\'{\i}sica Te\'orica, C-XI\hb
 Universidad Aut\'onoma de Madrid,\hb
 Canto Blanco,\hb
E-28049, Madrid, Spain.}\hb}
\smallskip
\centerline{\box9}
\bigskip

\setbox0=\vbox{\abstracttype{Abstract}
We improve, in the energy region between 
$\bar{K}K$ threshold and $\sim~1.4$~GeV, the energy-dependent 
phase shift analysis of  $\pi\pi$ scattering presented in a previous paper.
For the S0 wave we have included more data above 
$\bar{K}K$ threshold and we have taken into account systematically 
the elasticity data on the reaction $\pi\pi\to\bar{K}K$. We here made a coupled channel fit.
For the D0 wave we have considered information on low energy parameters, 
and imposed a better fit to the $f_2$ resonance. 
For both waves the expressions we now find are substantially more 
precise than the previous ones.
We also provide  slightly improved D2 and P waves, including
the estimated inelasticity  for the first, and a more flexible parametrization 
 between 1 and 1.42~GeV  for the second.
The accuracy of our amplitudes is now such 
that it requires a refinement of the Regge analysis, for $s^{1/2}\geq1.42$~GeV,
which we also carry out. We show that this more realistic input produces 
 $\pi\pi$ scattering amplitudes that satisfy better forward  dispersion relations, 
particularly for $\pi^0\pi^0$ scattering. 
}
\centerline{\box0}
\brochureendcover{Typeset with \physmatex}
\brochureb{\smallsc r. kami\'nski,  j. r. pel\'aez and f. j.  yndur\'ain}{\smallsc 
the pion-pion scattering amplitude. ii: improved analysis 
 above $\bar{K}K$ threshold}{1}

\brochuresection{1. Introduction}

\noindent
In a recent paper by two of us\ref{1} (JRP and FJY), that we will 
consistently denote by PY05, 
we have presented a set of fits to the data on 
$\pi\pi$ scattering phase shifts and inelasticities, and 
we also checked how well forward dispersion
relations  are satisfied by the different 
 $\pi\pi$ scattering phase shift analyses
(including our own). 
These various sets differ on the values of the S0 phase shifts below $\bar{K}K$ threshold.
We found that some of the most frequently used 
sets of phase shifts fail to satisfy  forward dispersion
relations and we then 
presented a consistent 
energy-dependent phase shift analysis of  
$\pi\pi$ scattering amplitudes that satisfies
well forward dispersion relations 
for energies below $\sim1\,\gev$. Above this energy, 
we found a certain mismatch 
between the real parts of the scattering amplitudes, calculated from phase
shifts and inelasticities,
and the result of the dispersive evaluations; 
particularly for $\pi^0\pi^0$ scattering.
This we attributed to imperfect experimental information in the 
region $1\gev\lsim s^{1/2}\lsim1.4\,\gev$. 

In the present paper we improve our analysis of the
S0 wave, the  D0 wave  and, to a lesser extent, the D2 and P 
waves\fnote{We will use consistently the 
self-explanatory notation S0, S2, P, D0, D2, F,\tdots\ for the 
$\pi\pi$ partial waves.} {\sl in the energy range around and above 
$\bar{K}K$ threshold}
  (for the D0 wave, we also slightly improve the low energy region). 
For the S0 wave 
we take
into account systematically 
the elasticity data from the reaction $\pi\pi\to\bar{K}K$; 
 for the D0 wave we
include information on low energy parameters,  
and we improve 
the fit to the $f_2(1270)$ resonance, to describe better its width
and inelasticity. 
These two parametrizations are more accurate than 
what we had in PY05; not only in that 
they include more data, but  also because they have {\sl smaller} errors.

A slight improvement for the P wave (using a more flexible parametrization)
 between $\bar{K}K$ threshold and 1.42~\gev\ is also 
presented and,
for the D2 wave, we improve on PY05 by including its
estimated inelasticity above $\sim1\,\gev$.

We have also found convenient to 
reconsider the Regge analysis, for the energy region above 1.42~\gev, 
particularly in view of the accuracy of the present parametrizations. 
This we do by taking into account more precise values for the 
intercepts $\alpha_\rho(0)$ and $\alpha_{P'}(0)$ than those used in PY05. 
Although the changes this induces  are very small, 
and indeed quite unnoticeable below 
1~\gev, the verification of the 
dispersion relation for 
exchange of isospin~1 above  $\bar{K}K$ threshold is sensitive to this Regge improvement.

We then also show that, with this more accurate input in the phase shift analysis, 
the forward $\pi^0\pi^0$  dispersion
relation 
is much better satisfied than with the amplitudes in PY05; 
particularly for energies above $1\;\gev$.
The $\pi^0\pi^+$  dispersion relation is also   improved, but only a little. 
Finally, the dispersion relation for exchange of isospin unity is 
practically unchanged below 1~\gev, and deteriorates slightly above.  
The new, improved parametrizations, therefore provide a very 
precise and reliable representation of
pion-pion  amplitudes at all energies: the average fulfillment of the dispersion relations 
is at the level of $1.05\,\sigma$, for energies below 0.93~\gev, and of $1.29\,\sigma$ 
for energies up to $1.42$~\gev.

\booksection{2. The S0 wave at high energy}

\noindent
In PY05 we  provided  fits to data for the S0 wave that satisfied
forward dispersion relations reasonably well below 0.925~\gev, 
as well as an improved parametrization constrained 
to satisfy forward dispersion relations below this energy and to fit data.
Here we will concentrate  on a parametrization at higher energies,
taking care to match it to the low energy one, 
which we do at 0.92 \gev.

The information on the S0 wave 
at high energy ($s^{1/2}>0.92\,\gev$) comes from
two sources: $\pi\pi$ scattering experiments\ref{2 - 6} and, 
above $\bar{K}K$ threshold, also from  
$\pi\pi\to\bar{K}K$ scattering.\ref{7} 
The second provides  reliable measurements of 
the elasticity parameter,\fnote{In the present paper
we refer to  $\eta$ as the {\sl elasticity}, or elasticity parameter. The 
inelasticity is $\sqrt{1-\eta^2}$.}
$\eta_0^{(0)}(s)$: since there are no isospin 2 waves 
in $\pi\pi\to\bar{K}K$ scattering, 
and the  $\pi\pi-\bar{K}K$ coupling is very weak for P and D0 waves,
 it follows 
that measurements of the differential cross section for 
 $\pi\pi\to\bar{K}K$ give directly the quantity
 $1-[\eta_0^{(0)}]^2$ with good accuracy, so long as the 
multipion cross section is small; see the discussion of this
below.

Below $\bar{K}K$ threshold we fit 
 data between 0.929~GeV and 0.970~\gev\ from Hyams et al.,\ref{2}
Protopopescu et al.\ref{3} and from 
Grayer et al.,\ref{4}
 as  composed in PY05:\ref{1}    

$$\eqalign{
\delta_0^{(0)}(0.929^2\,\gev^2)=&\,112.5\pm13\degrees;\quad
\delta_0^{(0)}(0.935^2\,\gev^2)=109\pm8\degrees;\cr
\delta_0^{(0)}(0.952^2\,\gev^2)=&\,126\pm16\degrees;\quad
\delta_0^{(0)}(0.965^2\,\gev^2)=134\pm14\degrees;\cr 
\delta_0^{(0)}(0.970^2\,\gev^2)=&\,141\pm18\degrees.\cr
}
\equn{(2.1a)}$$
As explained in PY05, these errors cover the {\sl systematic} uncertainties,
which are large. We also add the recent data of Kami\'nski et al.\ref{5}
and  we include in the fit the value 
 $$\delta_0^{(0)}(4m_K^2)=205\pm8\degrees
\equn{(2.1b)}$$ 
obtained in the constant K-matrix fit of Hyams et al.,\ref{2} which is compatible with 
the other 
data used here. Finally, we include  two values that follow from the low energy
analysis in PY05 
from the global data fit (i.e., before imposing 
forward dispersion relations, to avoid correlations with other waves),
$$\delta_0^{(0)}(0.900^2\;\gev^2)=101.0\pm3.7;\degrees,\quad
\delta_0^{(0)}(0.920^2\;\gev^2)=102.6\pm4\degrees.
\equn{(2.1c)}$$

To fit the data above $\bar{K}K$ threshold, we notice that analyses based on
 $\pi\pi$ scattering experiments only determine a combination
 of phase shift and
inelasticity and, indeed,  
 different results are obtained for the S0 wave in  the various analyses. 
For this wave we only fit data sets whose
 inelasticity is {\sl compatible} 
with what is  found in 
$\pi\pi\to\bar{K}K$ scattering,\ref{7} in the 
region $4m^2_K\leq s\lsim(1.25\,\gev)^2$. This includes
 the solution\fnote{$(---)$ is the preferred solution in the original reference.
Unfortunately, this reference only
 provided numbers for the statistical uncertainties. 
We add to these $5^{\rm o}$ as estimated
systematic error, in agreement with an  analysis similar to that of 
PY05.} $(---)$ of 
Hyams et al.,\ref{6} the data of Hyams et al.\ref{2} (or\fnote{The
 data of Grayer et al. in ref~3, and 
those of Hyams et al. in ref.~2 come from the same experiment.} of Grayer et
al.\ref{3}) and  the data of Kami\'nski et al.\ref{5}

For the elasticity parameter, $\eta_0^{(0)}$, we will improve 
on the analysis of PY05 by including more data\fnote{In all cases we add 
an estimated error of 0.04 to data that only give statistical errors.} (especially,
 $\pi\pi\to\bar{K}K$ data) and being more 
realistic in the parametrization. 
First of all, we remark that the modulus squared of the 
S0 amplitude for  $\pi\pi\to\bar{K}K$ scattering is proportional to 
$\tfrac{1}{4}(1-[\eta_0^{(0)}]^2)$, provided the two-channel 
approximation is valid. 
This is known to be the case experimentally for $s^{1/2}\lsim1.25\,\gev$ 
for such waves as has been measured, and will very likely be also true for our 
case (as we verify in Appendix~B). 
In this range, the  $\pi\pi\to\bar{K}K$ scattering experiments
 give the more
reliable  measurements of the parameter $\eta_0^{(0)}$. 
Therefore, in the region $s^{1/2}\lsim1.25\,\gev$ we fit 
 $\pi\pi\to\bar{K}K$ data\ref{7} and, 
among the $\pi\pi\to\pi\pi$ data sets,  only those whose inelasticity is
compatible with 
that from  $\pi\pi\to\bar{K}K$ below $\sim1.25\,\gev$. 
This includes the data sets of 
Hyams et al.\ref{2} (or Grayer et al.\ref{3});
  the data from ref.~6, solution $(- - -)$; and the data of
ref.~5. We however do {\sl not} include in the fits the
data of Protopopescu et al.,\ref{4} 
since they are quite incompatible with the   $\pi\pi\to\bar{K}K$
information.

A convenient way to fit phase shift and inelasticity is to use the K-matrix formalism.
This has the advantage over the  method of polynomial fits, used in PY05 
(see also Appendix~B here), that the relations that occur at threshold
between 
$\delta_0^{(0)}$ and  $\eta_0^{(0)}$, given in  Appendix~A [Eq.~(A.6)] 
are automatically fulfilled. The method, however,  presents the drawback that it is not possible
to take  into account the existence of other channels unless one introduces an 
excessive number of parameters. This is why we present an alternate polynomial fit in Appendix~B.
Fortunately, the fits given in Appendix~B   show that the contribution of such 
multiparticle channels is 
rather small; in fact, within the errors of the two channel fit
(this smallness is probably due to the fact that, because of its quantum numbers, the first
quasi-two body channel that contributes is  $\rho\rho$).
 So  we would expect that neglecting those other channels will not 
produce an excessive bias, being anyway covered by our uncertainties.

\topinsert{
\setbox0=\vbox{{\psfig{figure=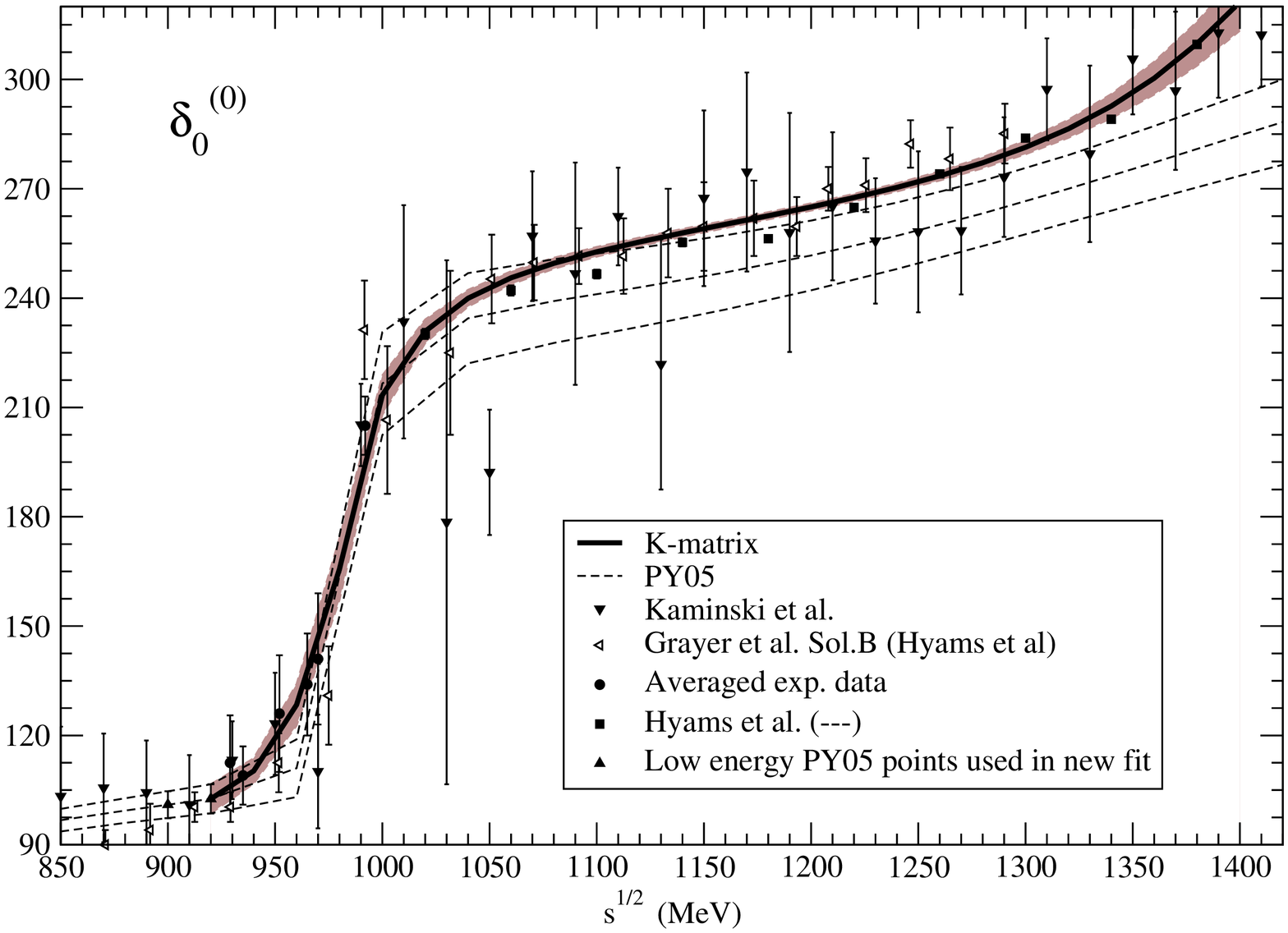,width=14truecm,angle=-90}}} 
\setbox6=\vbox{\hsize 15truecm\captiontype\figurasc{Figure 1A. }{
K-matrix fit to $\delta_0^{(0)}$ (solid line and dark area). 
Dotted lines: the fit in PY05.}\hb} 
\centerline{\otightboxit{\box0}}
\bigskip
\centerline{\box6}
\medskip
}\endinsert

To perform the fit, we consider  
$\delta_0^{(0)}$ and  $\eta_0^{(0)}$ to be given in terms of the 
K-matrix elements by the expressions (cf.  Appendix~A)
$$
\tan\delta_0^{(0)}(s)=\cases{\dfrac{k_1|k_2|\det {\bf K}+k_1K_{11}}{1+|k_2|K_{22}},
\quad s\leq 4m^2_K,\cr
\eqalign{&\,\dfrac{1}{2k_1[K_{11}+k_2^2K_{22}\det {\bf K}]}\Bigg\{
k^2_1K^2_{11}-k^2_2K^2_{22}+k_1^2k_2^2(\det{\bf K})^2-1\cr
+&\,
\sqrt{(k^2_1K^2_{11}+k^2_2K^2_{22}+k_1^2k_2^2(\det{\bf K})^2+1)^2-
4k_1^2k_2^2K^4_{12}}\;\Bigg\},\quad s\geq 4m^2_K} 
\cr}
\equn{(2.2)}$$
and
$$\eta_0^{(0)}(s)=\sqrt{\dfrac{(1+k_1k_2\det{\bf K})^2+(k_1K_{11}-k_2K_{22})^2}
{(1-k_1k_2\det{\bf K})^2+(k_1K_{11}+k_2K_{22})^2}},\quad s\geq 4m^2_K.
\equn{(2.3)}$$
 Then we 
 write a standard diadic expansion for $\bf K$, with a constant background 
(like, e.g.,
in Hyams et al.\ref{2}):
$$K_{ij}(s)=\dfrac{\mu\alpha_i \alpha_j}{M_1^2-s}+
\dfrac{\mu\beta_i\beta_j}{M_2^2-s}+\dfrac{1}{\mu}\gamma_{ij}
\equn{(2.4a)}$$
and $\mu$ is a mass scale, that we take $\mu=1\,\gev$.
The powers of $\mu$ have been arranged so that the $\alpha_i$, $\beta_i$, $\gamma_{ij}$ are
dimensionless;  they are also assumed to be 
 constant.
The pole at $M_1^2$ simulates the left hand cut of $\bf K$, and 
the pole at $M_2^2$ 
is connected with the phase shift crossing $270\degrees$ around 1.3~\gev;
both poles are necessary to get a good fit.

\topinsert{
\setbox0=\vbox{{\psfig{figure=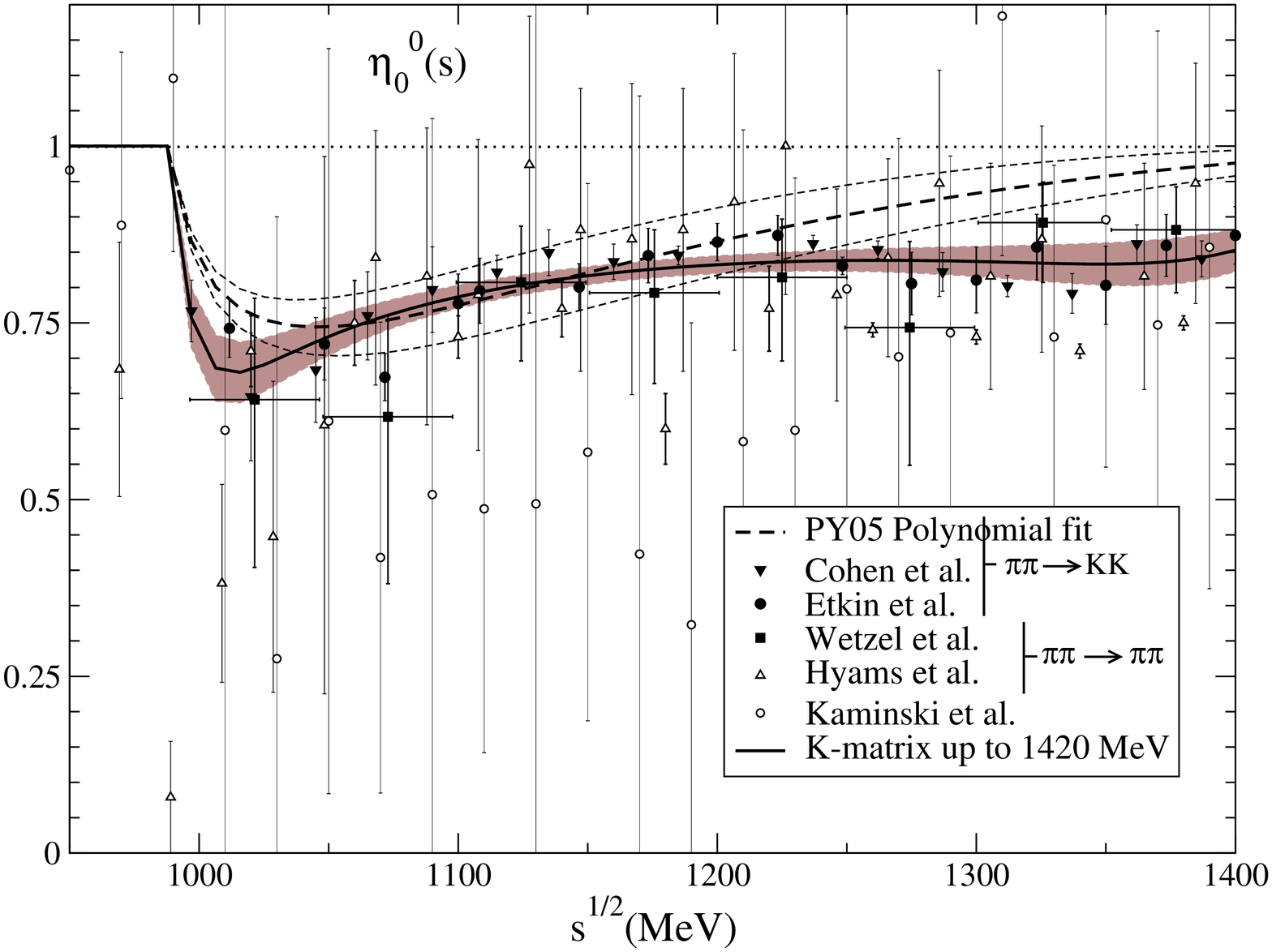,width=12.truecm,angle=-90}}} 
\setbox6=\vbox{\hsize 15truecm\captiontype\figurasc{Figure 1B. }
{K-matrix fit to $\eta_0^{(0)}$ using (2.4), with 
error given by the shaded area. 
The dotted lines represent the central values and error limits of the old fit in PY05.}\hb
\phantom{x}\hb
} 
\centerline{\otightboxit{\box0}}
\medskip
\centerline{\box6}
\medskip
}\endinsert

We fit simultaneously all data, 
above as well as below $\bar{K}K$ threshold for the phase shift. For $\eta_0^{(0)}$, 
we also fit all data, $\pi\pi\to\pi\pi$ and $\pi\pi\to\bar{K}K$, over the 
whole range, which is justified since we are neglecting 
other inelastic channels. We require perfect matching  with the lower energy determination 
of the phase shift at 0.932~\gev, as obtained in PY05. 
We find a \chidof=0.6 and the
 values of the parameters
are
$$\eqalign{\alpha_1=&\,0.727\pm0.014,\quad \alpha_2=0.19\pm0.04,\quad \beta_1=1.01\pm0.08;\quad
\beta_2=1.29\pm0.03,\cr M_1=&\,910.5\pm7\;\mev,\quad M_2=1324\pm6\;\mev;\cr
\gamma_{11}=&\,2.87\pm0.17,\quad \gamma_{12}=1.93\pm0.18,\quad \gamma_{22}=-6.44\pm0.17;\cr
\delta_0^{(0)}&\,((0.92\,\gev)^2)=103.55\pm4.6\degrees.
\cr}
\equn{(2.4b)}$$
Note that $M_1$  indeed lies near the beginning of the left hand cut for
 $\bar{K}K\to\pi\pi$ scattering, located at 952~\mev.

The parameters in (2.4b) are strongly correlated. In fact, we have verified that there exists 
a wide set of minima, with very different values of the parameters. 
This is not surprising, since we do not have sufficiently many observables to 
determine the three  $K_{ij}$ on an energy 
independent basis.
 Nevertheless, 
the corresponding values of $\delta_0^{(0)}$ and $\eta_0^{(0)}$  vary
very little in all
these minima, so that (2.4b) can be considered a faithful  
representation of the S0 wave for $\pi\pi$ scattering, albeit very likely with
somewhat underestimated 
errors due to our neglecting other channels, like $\pi\pi\to4\pi$.
The corresponding phase shift and elasticity parameter are shown in Figs.~1A,~1B.
Both phase shift and elasticity clearly improve what we had in PY05.

It is mainly because of the use of the phase shift in (2.4a) and, above all, 
the smaller inelasticity driven by $\pi\pi\to\bar{K}K$ data, that we find
a  substantial improvement in forward $\pi^0\pi^0$ dispersion relations above 
1~\gev\ (see below), as 
already remarked in PY05.

\booksection{3. The improved D0, D2 waves}
\vskip-0.5truecm
\booksubsection{3.1. The D0 wave}

\noindent
The experimental data on the D0 wave are of poor quality; different experiments give 
very incompatible results, particularly below 0.93~\gev. Above 1.1~\gev, and 
although the data of refs.~2,~6 (but not ref.~3) 
are  compatible, 
it is better to use directly the Particle Data Table's information\ref{8} on 
the $f_2$ resonance, which summarizes the existing experimental data.
\goodbreak 
The reliable information on this wave is then of three kinds.
Firstly, in the range around 
1.27~\gev, we have the referred very precise measurements of the $f_2(1270)$ resonance
parameters, which give\ref{8} a mass 
$M_{f_2}=1275.4\pm1.2\,\mev$, 
 a width $\gammav_{f_2}=185.1\pm3.4\,\mev$ and a $\pi\pi$ branching ratio of $84.7\pm2.4\%$.
Secondly,  the Froissart--Gribov representation allows 
an accurate determination of the scattering length,  $a_2^{(0)}$, and 
effective range parameter, 
$b_2^{(0)}$, as shown in PY05;\fnote{The values given below in (3.1) are those obtained in 
PY05, with the old parametrizations. 
We have verified that they do {\sl not} change, within the accuracy 
of (3.1), if recalculating  the Froissart--Gribov representation with 
the parametrizations in the present paper. This  of course occurs because the 
new parametrizations only change the amplitudes significantly above 1~\gev, a region to which 
 $a_2^{(0)}$ and  
$b_2^{(0)}$ are almost not sensitive.} one
finds\ref{1}
$$a_2^{(0)}=(18.7\pm0.4)\times10^{-4}\,M_{\pi}^{-5},\quad
b_2^{(0)}=(-4.2\pm0.3)\times10^{-4}\,M_{\pi}^{-7}.
\equn{(3.1)}$$
This helps us to fix the phase shift at low energy.
And thirdly, we have the 1973 data of 
Hyams et al.\ref{2}, Protopopescu et al.,\ref{4} 
and solution $(- - -)$ of Hyams et al.\ref{6}
in the range $0.935\,\gev\leq s^{1/2}\leq1.1\,\gev$, which are
reasonably  compatible among themselves; see \fig~2.
We denote them by, respectively, H73, P and H$(- - -)$. 
The data we include thus are
$$
\matrix{E,\;{\rm in}\;\gev&\cot\delta_2^{(0)}&{\rm source}\cr
0.935&10.4\pm2&{\rm P}\cr
0.950&10.2\pm5.5&{\rm H73}\cr
0.965&7.3\pm2&{\rm P}\cr
0.970&4.0\pm2.2&{\rm H73}\cr
0.990&4.8\pm1.4&{\rm H73}\cr}
\equn{(3.2a)}$$
and, above $\bar{K}K$ threshold,
$$\eqalign{
\matrix{E,\;{\rm in}\;\gev&\cot\delta_2^{(0)}&{\rm source}\cr
1.00&5.1\pm2&{\rm P}\cr
1.01&3.6\pm0.8&{\rm H73}\cr
1.02&3.6\pm0.8&{\rm H}\,(- - -)\cr
1.03&3.0\pm0.6&{\rm H73}\cr
1.04&3.3\pm0.8&{\rm P}\cr}\qquad\qquad
\matrix{\cr
E,\;{\rm in}\;\gev&\cot\delta_2^{(0)}&{\rm source}\cr
1.05&3.8\pm0.9&{\rm H73}\cr
1.06&2.8\pm0.8&{\rm H}\,(- - -)\cr
1.07&2.5\pm0.4&{\rm H73}\cr
1.09&2.7\pm0.45&{\rm H73}\cr
1.10&2.1\pm0.8&{\rm H}\,(- - -).\cr
\cr}\cr}
\equn{(3.2b)}$$
A few words must be said about the errors in (3.2).
Since  H$(- - -)$ do not give errors, we take them as equal to those of P. 
We also multiplied {\sl all} errors by a factor 2, to take into account 
the estimated systematic
errors (for e.g. P, estimated as the difference between the fits XIII and VI 
in ref.~3).

\topinsert{
\setbox0=\vbox{{\psfig{figure=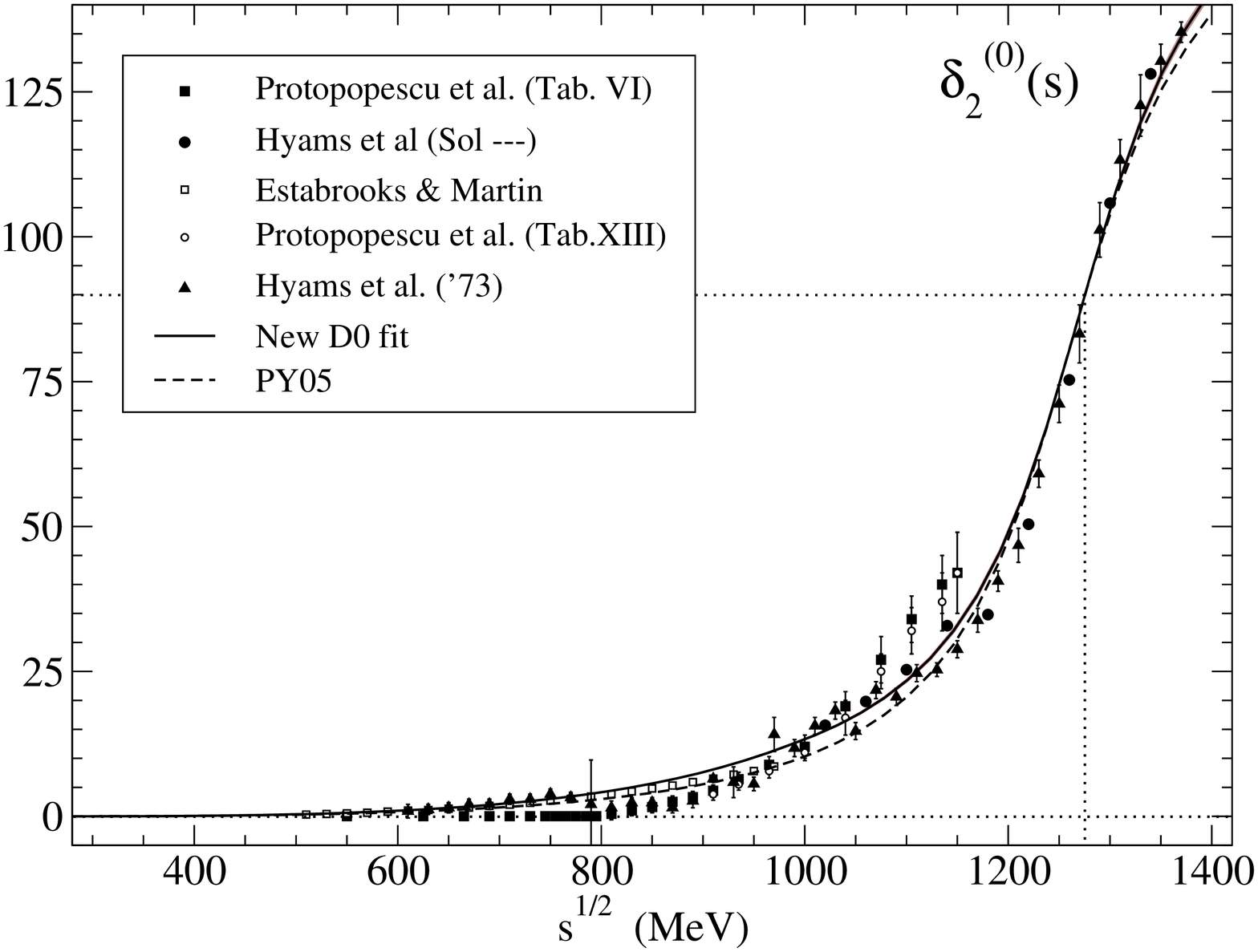,width=11.2truecm,angle=-90}}}  
\setbox6=\vbox{\hsize 14.0truecm\noindent\petit\figurasc{Figure 2. }{The 
D0 phase shift as determined here (continuous line; the error 
is like the thickness of the line) and that from
  PY05 
(broken line). 
 The experimental data
points  are also shown. Note that the high energy fit is 
tightly
constrained by the $f_2(1270)$ 
mass and width.
}} 
\centerline{\otightboxit{\box0}}
\bigskip
\centerline{\box6}
\bigskip}\endinsert

We  present the details of the fits. To 
take into account the analyticity structure, we
fit with different expressions for energies below and above $\bar{K}K$ threshold, 
requiring however exact matching at $s=4m^2_K$.
Below $\bar{K}K$ threshold we take into account the existence of nonnegligible 
inelasticity above 1.05~\gev, which is near the $\omega\pi$ or $\rho\pi\pi$ 
thresholds, by choosing a conformal variable $w$ appropriate to a plane cut 
for $s>(1.05\,\gev)^2$. 
So we write
$$\eqalign{
\cot\delta_2^{(0)}(s)=&\,\dfrac{s^{1/2}}{2k^5}\left(M^2_{f_2}-s\right)M^2_\pi
\Big\{B_0+B_1w\Big\},\quad  s< 4m_K^2;\cr
w=&\,\dfrac{\sqrt{s}-\sqrt{\hat{s}-s}}{\sqrt{s}+\sqrt{\hat{s}-s}},\quad
\hat{s}^{1/2}=1.05\;\gev.\cr
}
\equn{(3.3a)}$$
The mass of the $f_2$ we fix at $M_{f_2}=1275.4\,\mev$; no error is taken for 
this quantity, since it is
negligibly small (1.2~\mev) when compared with the other errors.
We fit the values of  $a_2^{(0)}$ and  
$b_2^{(0)}$ given in (3.1) and the data in (3.2a). We find 
 the values of the parameters
$$B_{0}=12.47\pm0.12;\quad B_{1}=10.12\pm0.16.
\equn{(3.3b)}$$ 
We note that the series shows good convergence.

Above  $\bar{K}K$ threshold we use the following formula for the phase shift:
$$\eqalign{
\cot\delta_2^{(0)}(s)=&\,\dfrac{s^{1/2}}{2k^5}\left(M^2_{f_2}-s\right)M^2_\pi
\Big\{B_{h0}+B_{h1}w\Big\},\quad  s> 4m_K^2;\cr
w=&\,\dfrac{\sqrt{s}-\sqrt{{s_h}-s}}{\sqrt{s}+\sqrt{s_h-s}};\quad
s_h^{1/2}=1.45\;\gev.\cr
}
\equn{(3.4a)}$$
This neglects inelasticity below 1.45~\gev, which is approximately the $\rho\rho$ threshold; 
inelasticity will be added by hand, see below.
We then fit the values for the width of $f_2(1270)$,
  as given above, and the  set of data in (3.2b). 
We get 
$$B_{h0}=18.77\pm0.16;\quad B_{h1}=43.7\pm1.8.
\equn{(3.4b)}$$ 
As stated, we have required exact matching of high and low 
energy  at $\bar{K}K$ threshold, 
where our fits give
$\cot\delta_2^{(0)}(4m_K^2)=4.42\pm0.04$.
This matching implies that there is a relation among the four 
$B_i$s, so there are in effect only three free parameters.
The overall chi-squared of the fit is very good,  $\chidof=9.8/(18-3)$.

\topinsert{
\setbox0=\vbox{\psfig{figure=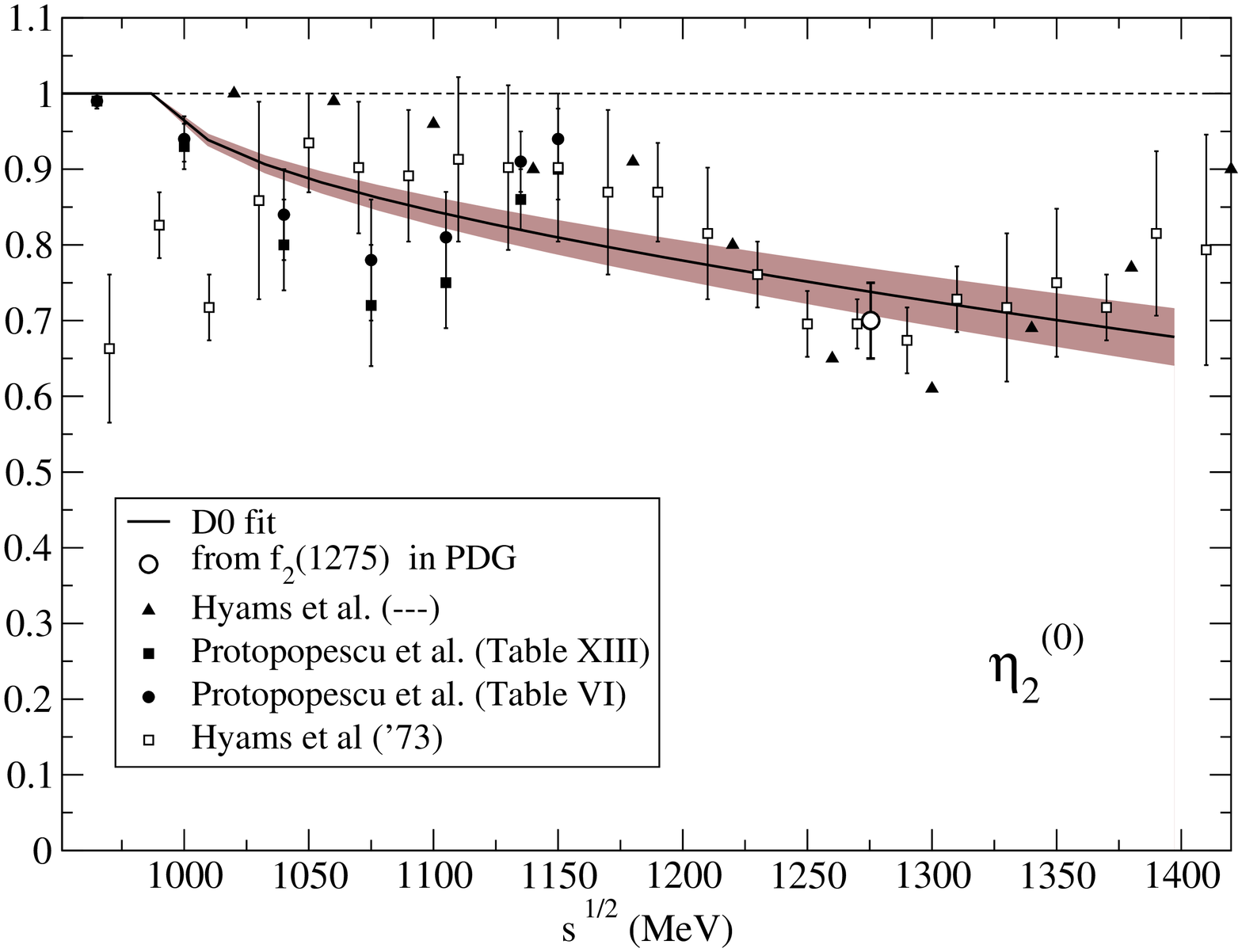,width=9.5truecm,angle=-90}}  
\setbox6=\vbox{\hsize 6truecm\noindent\petit\figurasc{Figure 3. }{
Fit to $\eta_2^{(0)}$ (continuous line and dark area that covers the uncertainty),
from PY05. Data from refs.~2,~3,~4,~6.
The  elasticity on the $f_2(1270)$, from 
 the PDT,\ref{8} is also shown (large white dot).
}} 
\line{\otightboxit{\box0}\hfil\box6}
\bigskip}\endinsert

We note that, although not included in the fit, our new D0 
phase shift fits better than the old PY05 one the data points of Hyams et al.,\ref{6}
solution $(- - -)$, above the $f_2$ resonance: see \fig~2.

The data for the inelasticity are not sufficiently good 
to  improve significantly the fit in PY05; so we simply write, as in ref.~1,
$$\eqalign{
\eta_2^{(0)}(s)=&\,\cases{1,\qquad  s< 4m_K^2,\;
\cr
1-\epsilon\,\dfrac{k_2(s)}{k_2(M^2_{f_2})},\quad
 s> 4m_K^2;\quad\epsilon=0.262\pm0.030,\quad
 k_2=\sqrt{s/4-m^2_K}.\cr} \cr
}
\equn{(3.4c)}$$
This probably only provides a fit to elasticity parameter on the average, 
but we have not been able to find a clear improvement on this.
The corresponding elasticity parameter is shown\fnote{Although 
there is nothing new in this fit, 
we show the picture because we had not shown it in PY05.} in \fig~3.

Two important properties of the new fit
are that it reproduces better than the one in PY05 
the width and inelasticity of the $f_2$ resonance, 
which is the 
more salient feature of the D0 wave, and that 
it is  more precise than what we had in PY05.
This improvement of the D0 wave, although 
it does not give a phase shift very different from that 
in PY05,  also contributes a nonnegligible amount to the
 improved fulfillment of the 
$\pi^0\pi^0$ dispersion relations.

\booksubsection{3.2. The D2 wave}

\noindent
In PY05 we fitted the D2 wave with a single parametrization
 over the whole energy range up to 
1.42~\gev, and neglected inelasticity. 
We wrote
$$\cot\delta_2^{(2)}(s)=
\dfrac{s^{1/2}}{2k^5}\,\Big\{B_0+B_1 w(s)+B_2 w(s)^2\Big\}\,
\dfrac{{M_\pi}^4 s}{4({M_\pi}^2+\deltav^2)-s}
\equn{(3.5a)}$$
with $\deltav$ a free parameter fixing the zero 
of the phase shift near threshold,  and
$$w(s)=\dfrac{\sqrt{s}-\sqrt{s_0-s}}{\sqrt{s}+\sqrt{s_0-s}},\quad
 s_0^{1/2}=1450\,\mev.$$
Since  the data  on this wave are not accurate  we
 included extra information. To be precise, we 
incorporated in the fit the value of  
 the scattering length  that follows
 from the Froissart--Gribov representation (PY05), 
$$a_2^{(2)}=(2.78\pm0.37)\times10^{-4}\,M_{\pi}^{-5},
$$
but {\sl not} that of the effective range parameter,
$$
\quad b_2^{(2)}=(-3.89\pm0.28)\times10^{-4}\,{M_\pi}^{-7}.
$$
We  got a mediocre fit, $\chidof=71/(25-3)$, and the values of the parameters
were
$$B_0=(2.4\pm0.3)\times10^3,\quad B_1=(7.8\pm0.8)\times10^3,\quad
 B_2=(23.7\pm3.8)\times10^3,\quad
\deltav=196\pm20\,\mev.
\equn{(3.5b)}$$
The corresponding numbers for the scattering length and  for the effective range
parameter 
$b_2^{(2)}$ that follow from this are 
$$a_2^{(2)}=(2.5\pm0.9)\times10^{-4}\,{M_\pi}^{-5};\quad
b_2^{(2)}=(-2.7\pm0.8)\times10^{-4}\,{M_\pi}^{-7}.
$$
The last is a bit away from what one has from the Froissart--Gribov representation, 
but still is compatible at the $2\,\sigma$ level.
The low quality of the fit may be traced to the fact that 
the various data sets are not very compatible among themselves. 
Therefore, there is no chance to improve the fit 
as we did for the D0 wave (where we had the very precise data on the $f_2$ resonance). 
We here merely improve the treatment of this wave 
by including  the inelasticity by hand.

To get an estimate of the inelasticity we have two possible methods: 
we can take the inelasticity 
to be similar to that of the D0 wave; or we can make a model 
calculation. For example, that of ref.~9, in which the authors 
assume inelasticity to go via rho intermediate states,
fixing the coupling parameters to reproduce the properties of the 
better known waves.  Both methods yield negligible inelasticity 
below $\rho\pi\pi$ threshold, and  something around
5\% inelasticity at the highest energy considered, 
1.42~\gev.
For the elasticity parameter we thus simply write, 
 above $1.05\,\gev$,
$$\eta_2^{(2)}(s)=1-\epsilon (1-\hat{s}/s)^3,\quad \hat{s}^{1/2}=1.05\;\gev,\quad
\epsilon=0.2\pm0.2;
\equn{(3.5c)}$$
this is negligible up to $1.25\,\gev$ and,
above that, covers both what was estimated in
ref.~9, and the fact that 
experiments fail to detect inelasticity.

\topinsert{
\setbox0=\vbox{{\psfig{figure=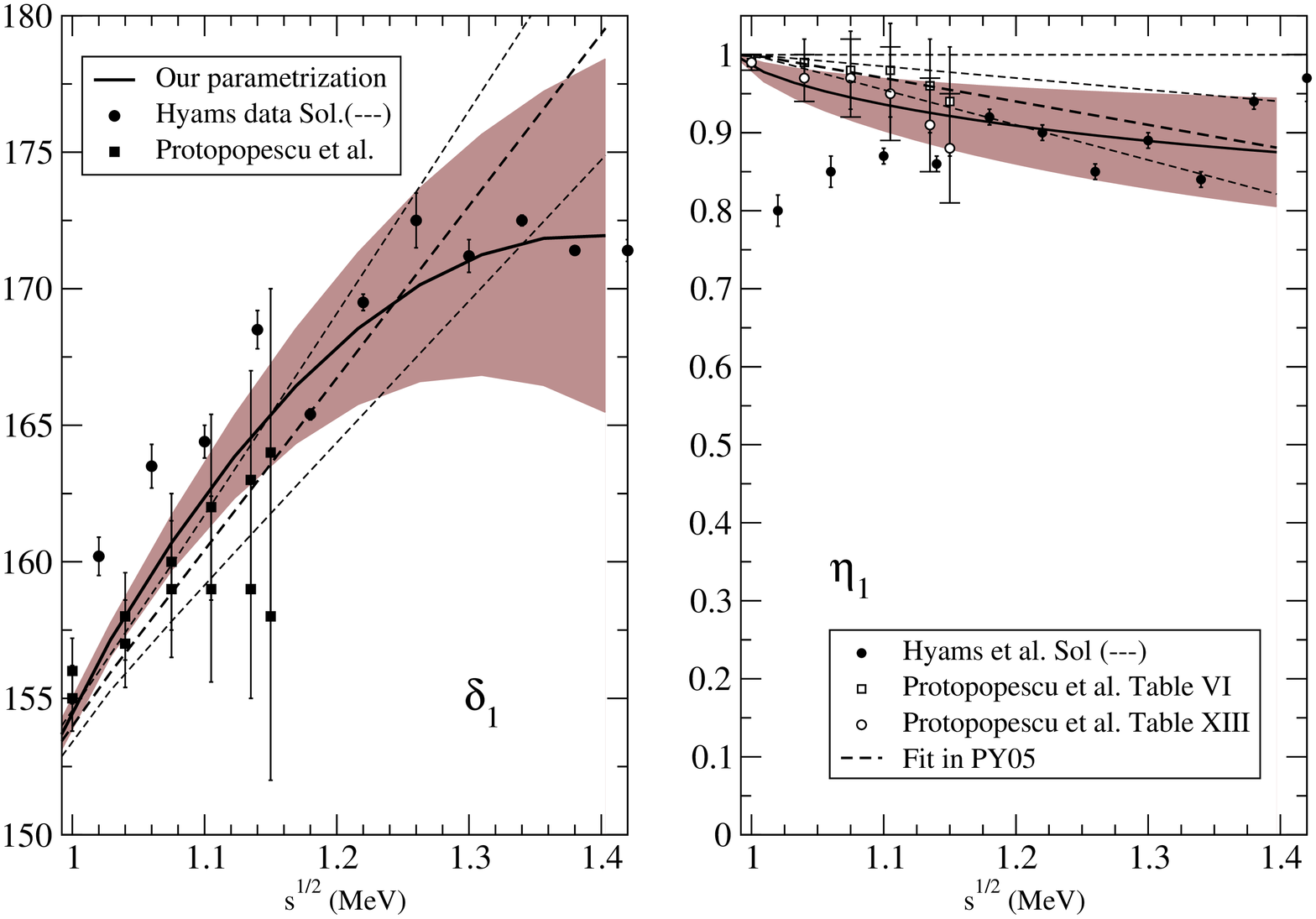,width=13.1truecm,angle=-90}}} 
\setbox6=\vbox{\hsize 15truecm\captiontype\figurasc{Figure 4. }{The fit
 to the P wave above $\bar{K}K$ threshold 
 (continuous line and dark area that covers the uncertainty), 
 with  data from solution $(---)$ 
of Hyams et al.\ref{6} and of Protopopescu et al.\ref{4} 
Note that the errors shown for the data are only the {\sl statistical} errors; 
systematic errors, estimated as in the text, about double them.
The broken lines
 are the phase shift and elasticity parameter of
PY05.
The effect of the $\phi(1020)$ resonance is not shown in this figure.}\hb} 
\centerline{\otightboxit{\box0}}
\bigskip
\centerline{\box6}
\medskip
}\endinsert
\booksection{4. The improved P
wave between $\bar K K$ threshold and 1.42~\gev}

\noindent We next fit the P wave above 
$2m_K\simeq0.992\,\gev$,  incorporating in
the  fit  the data from   solution $(- - -)$ of Hyams et al.,\ref{6} 
besides  the data from  Protopopescu et al.\ref{4} 
(the last is the  one more compatible with what one finds from the pion form factor).
We have added estimated  errors of $2\degrees$ to the phase shift and 0.04 to the
elasticity parameter for the data of solution $(- - -)$ in Hyams et
al.,\ref{6} since no errors are provided in this reference.
 We now use one more parameter both 
for the phase shift and for the elasticity parameter than what we had in PY05, writing 
$$\eqalign{\delta_1(s)=&\,\lambda_0+\lambda_1(\sqrt{s/4m^2_K}-1)+
\lambda_2(\sqrt{s/4m^2_K}-1)^2,\cr 
\eta_1(s)=&\,1-\epsilon_1\sqrt{1-4m^2_K/{s}}-\epsilon_2(1-4m^2_K/{s});
 \quad  s> 4m_K^2.\cr
}
\equn{(4.1)}$$ 
The phase at the low energy edge,
$\delta_1(0.992^2\,\gev^2)=153.5\pm0.6\degrees$, 
is obtained from the fit to the form factor of the pion 
(ref.~9; see also ref.~1).
 This fixes the vale of $\lambda_0$.

The fits are reasonable;   we get $\chidof=0.6$ for the phase and   $\chidof=1.1$ 
for the elasticity. 
 We find the parameters
$$\eqalign{\lambda_0=&\,
2.687\pm0.008,\quad\lambda_1=1.57\pm0.18,\quad \lambda_2=-1.96\pm0.49;\cr
\epsilon_1=&\,0.10\pm0.06,\quad \epsilon_2=0.11\pm0.11.\cr
}\equn{(4.2)}$$
The only noticeable differences with the fit in PY05 is that the  
 phase shift and elasticity parameter are now less rigid, 
that 
we match the low and high energy expressions at $\bar{K}K$ threshold,
and that the {\sl
inelasticity} is now somewhat larger than what we had in PY05.
This improved solution, together with that in 
 PY05, are shown in \fig~4.

Another matter is  the
contribution of  the $\phi(1020)$  resonance. 
This can be included in the standard way, by adding to the P wave 
a resonant piece
$$\hat{f}_1(s)\to\hat{f}_1(s)+\hat{f}_1^{\phi}(s),
\equn{(4.3a)}$$
where $\hat{f}_1(s)$ is normalized so that, in the elastic case,
$$\hat{f}_1=\sin\delta_1\ee^{\ii \delta_1}$$
and
$$ \hat{f}_1^\phi(s)=
\dfrac{\dfrac{M_\phi}{s^{1/2}}\,\left[\dfrac{k_2}{k_2(M^2_\phi)}\right]^3\,
M_\phi\gammav_\phi}
{M^2_\phi-s-\ii\dfrac{M_\phi}{s^{1/2}}\,\left[\dfrac{k_2}{k_2(M^2_\phi)}\right]^3
M_\phi\gammav_\phi}\, B_{\pi\pi}, \quad s\geq4m^2_K;
\equn{(4.3b)}$$
here $k_2=\sqrt{s/4-m_K^2}$, and the width and $\pi\pi$ branching ratio of the 
$\phi(1020)$ resonance are $\gammav_\phi=4.26\pm0.05\,\mev$ and  
$B_{\pi\pi}=(7.3\pm1.3)\times10^{-5}$.
Something similar could be done for the contribution of the $\omega$.

The influence of these resonances is totally 
negligible and,
in fact, we will {\sl not} include them in our calculations of dispersion relations 
below.

\booksection{5. Improvement of the Regge input}

\noindent
To evaluate the dispersion relations we need an estimate for the high 
energy ($s^{1/2}\geq1.42\,\gev$) scattering amplitudes. This is furnished by 
the Regge model. We have here three amplitudes, one for each of 
the  exchange of isospin 0, 1 and
2.  We first 
take, for these Regge amplitudes,  the results of the fits in ref.~11; see also PY05,
Appendix~B. Then  we 
will consider improvement of the 
Regge parameters.

\goodbreak

The expressions for the amplitudes for exchange of isospin 1 and 0 are, respectively, 
$$\imag F^{(I_t=1)}(s,0)\simeqsub_{s\to\infty}
\beta_\rho(0)(s/\hat{s})^{\alpha_\rho(0)},
\quad
s\geq(1.42\;{\gev})^2,
\equn{(5.1a)}$$ 
and
$$\eqalign{ 
\imag F^{(I_t=0)}_{\pi\pi}(s,0)&\,\simeqsub_{s\to\infty}
P(s,0)+P'(s,0),\quad
s\geq(1.42\;{\gev})^2;\cr
P(s,0)=&\,\beta_P\,(s/\hat{s})^{\alpha_P(0)},\quad
P'(s,0)=\beta_{P'}\,(s/\hat{s})^{\alpha_{P'}(0)}.
\cr
}
\equn{(5.1b)}$$
In both expressions
 $\hat{s}=1\,\gev^2$. 
The values of the parameters are (ref.~11 and ref.~1, Appendix~B) 
$$\beta_\rho(0)=1.02\pm0.11;\quad\alpha_\rho(0)=0.52\pm0.02
\equn{(5.2)}$$ 
and
$$\beta_P=2.54\pm0.03,\quad
\beta_{P'}=1.05\pm0.02;\quad\alpha_{P'}(0)=\alpha_\rho(0),\quad\alpha_P(0)=1.
\equn{(5.3)}$$
For exchange of isospin 2, which is very small, we also take 
the amplitude of ref.~11:
$$\imag F^{(I_t=2)}(s,0)\simeqsub_{s\to\infty}
\beta_2(s/\hat{s})^{2\alpha_\rho(0)-1},\quad\beta_2=0.2\pm0.2; \quad
s\geq(1.42\;{\gev})^2.
\equn{(5.4)}$$

The first two, however, will now be improved: 
 as we have seen in 
previous sections, the precision of our new parametrizations in the intermediate
energy  range ($\sim1$ to $\sim1.4$~\gev) is such that one is sensitive 
to  small details of the Regge amplitudes; so,  
 it is convenient to re-assess the derivation of 
the values for the Regge parameters  in Eqs.~(5.2) to (5.3).

The expressions (5.2), (5.3) were obtained in ref.~11 and PY05 as follows. 
We fixed
${\alpha}_{\rho}(0)$ as the average between what is found in deep inelastic scattering,\ref{12}
${\alpha}_{\rho}=0.48$, and 
in the analysis of hadron collisions by Rarita et al.,\ref{13} who get 
$\alpha_\rho=0.56$. We also imposed 
degeneracy, so that the intercept of 
$\rho$ and $P'$ were forced to be the same.
We then fitted
 experimental $\pi\pi$ cross sections, which 
gives $\beta_\rho(0)=1.0\pm0.3$,   and improved this result 
demanding fulfillment of a crossing sum rule. 
For isospin zero exchange, the expression (5.3) was obtained requiring 
simultaneous fits to $\pi\pi$, $\pi N$ and $NN$ 
data, using factorization, fixing the intercept of the $P'$ 
to 0.52 (as already stated).

However, more complete fits\ref{14} than that of Rarita et al.\ref{13} 
have been performed in the last years; especially, 
for the rho trajectory, individual data on $pp$, $\bar{p}p$ and $np$ 
have been included in the fits, which permits improvement of the determination of 
the rho parameters using factorization.
These fits, 
 in particular, allowed a relaxation of the exact degeneracy 
condition $\alpha_\rho(0)=\alpha_{P'}(0)$, and
yield central values for the rho intercept $\alpha_\rho=0.46$, more in agreement with the result 
from deep inelastic scattering. For  $\alpha_{P'}$ one finds a value higher than 
for the rho intercept:  
$\alpha_{P'}=0.54$.

We may then repeat the analysis of ref.~11, but fixing now 
the intercepts of rho and $P'$ trajectories to the likely more 
precise values
$$\alpha_\rho(0)=0.46\pm0.02,\quad\alpha_{P'}(0)=0.54\pm0.02,
\equn{(5.5a)}$$
with conservative errors.
We also here improve the error estimate for the rho residue $\beta_\rho$
with the crossing sum rule, as we did in PY05 to get (5.2). 
This sum rule we calculate  using 
the new phase shifts and inelasticities we have evaluated in the present paper.
We then find,
$$\beta_\rho=1.22\pm0.14,\quad\beta_P=2.54\pm0.04,\quad\beta_{P'}=0.83\pm0.05.
\equn{(5.5b)}$$ 
The errors are slightly larger now, which is due to the fact that we do not impose 
the exact degeneracy relation
$\alpha_\rho(0)=\alpha_{P'}(0)$. 
For the amplitude with exchange of isospin 2, we still keep (5.4) since no new 
information is available.

The difference between what we have now, (5.5), and what was used in PY05 is much
smaller than what would appear at first sight; in fact, because the 
$\alpha(0)$ and $\beta$ are strongly correlated, the 
changes in one quantity are compensated by those in the other:
the 
{\sl amplitudes} described by (5.5) and (5.2,~3) are very similar  
in the energy region of interest (cf. \fig~5). 
However, these amplitudes differ in some details. 
So, the rho amplitude described by (5.5) 
is tilted with respect to that given by (5.2): 
the amplitude
described by (5.5) is slightly larger than that described by
  (5.3) below $\sim5~\gev$, where they
cross over, and is larger above this energy. 
Likewise, for exchange of isospin zero (5.5) gives a smaller 
amplitude at low energy, which then crosses over the amplitude 
given by (5.3) at higher energy.

\topinsert{
\setbox2=\vbox{\psfig{figure=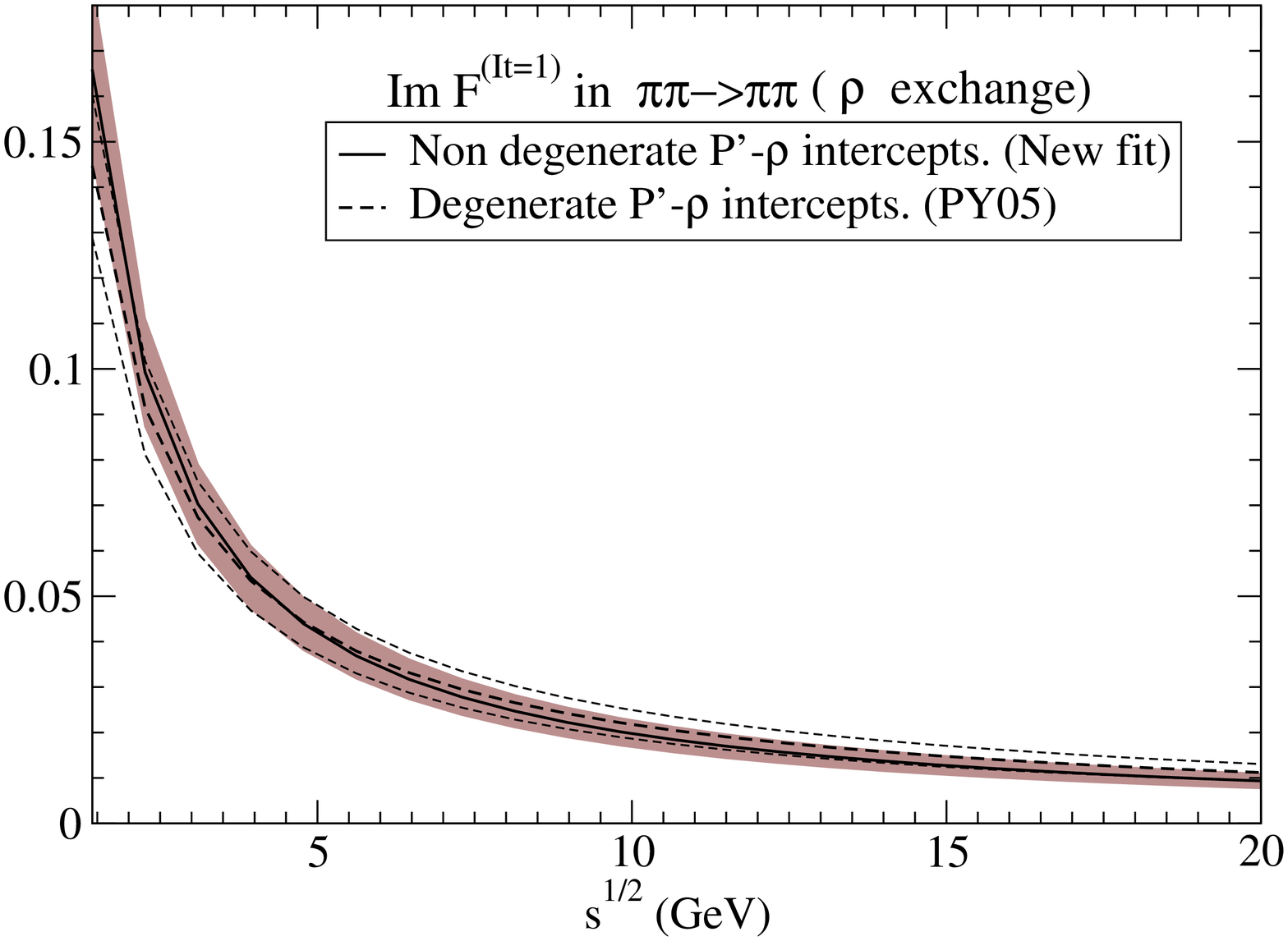,width=11.2truecm,angle=-90}} 
\setbox3=\vbox{\hsize 4.8truecm\captiontype\figurasc{Figure 5A. }{\hb
The 
scattering amplitude $\imag F^{(I_t=1)}(s,0)$ as described by 
(5.2), broken line, and (5.5), solid line with error included (gray band).\hb
\phantom{K}}} 
\line{\otightboxit{\box2}\hfil\box3}
\bigskip
\setbox4=\vbox{\psfig{figure=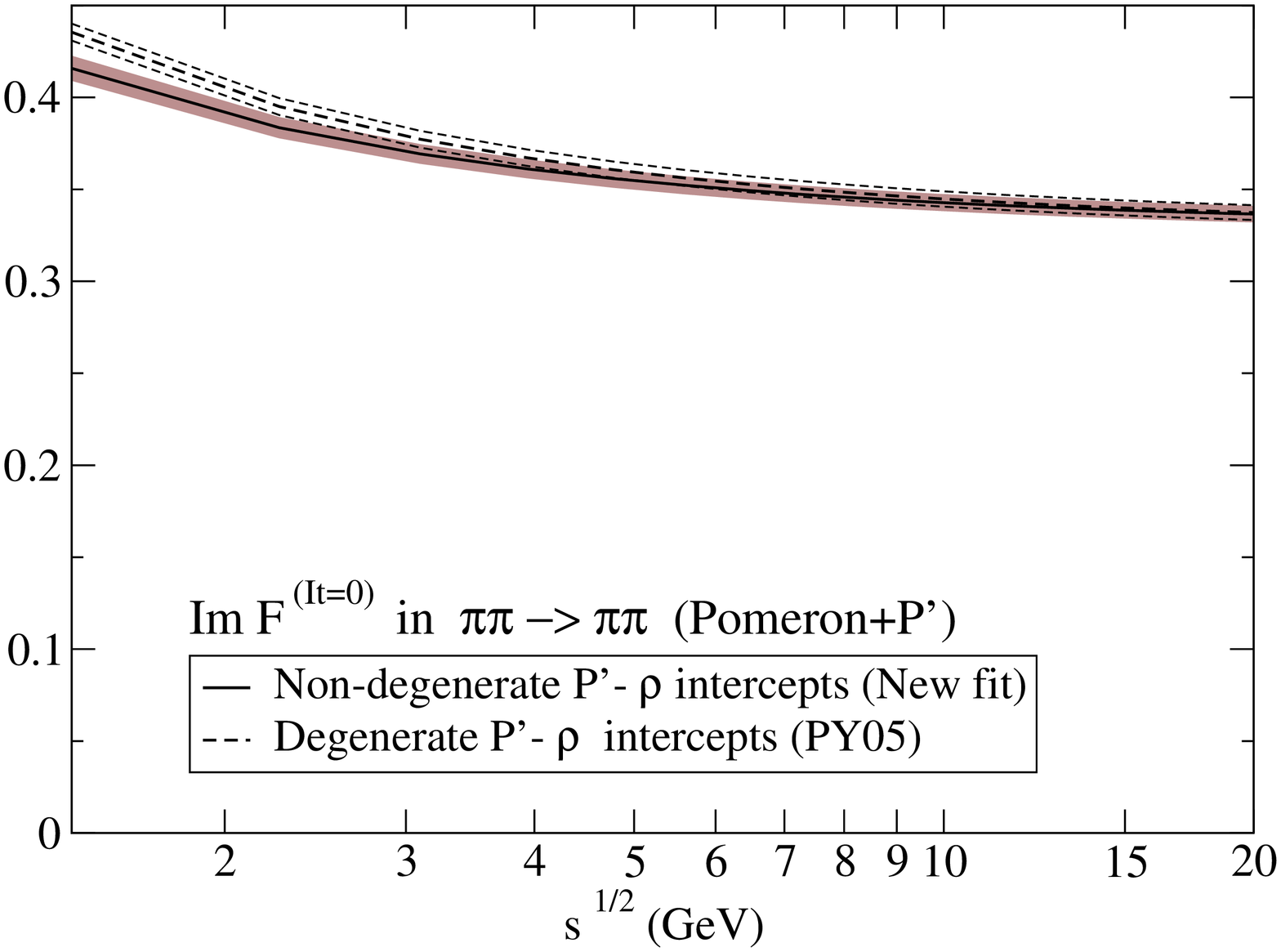,width=11.2truecm,angle=-90}} 
\setbox5=\vbox{\hsize 4.8truecm\captiontype\figurasc{Figure 5B. }{\hb
The 
scattering amplitude $\imag F^{(I_t=0)}(s,0)$ as described by 
(5.3), broken line, and (5.5), solid line with error included (gray band).\hb
\phantom{K}}} 
\line{\otightboxit{\box4}\hfil\box5}
\bigskip
}\endinsert

As just stated, these changes induced by using (5.5) do almost 
compensate each other and, indeed, they have only a minute effect 
in dispersion relations below 1~\gev. At the level of precision
attained by our  parametrizations in the region above 1~\gev, however,
the dispersion relations are  sensitive to the  
details of the Regge behaviour; 
because of this, we  will evaluate the dispersion relations 
 with both (5.2,~4) and with (5.5).

A last question related to the high energy, $s^{1/2}\geq1.42\,\gev$, input is 
the matching of the Regge amplitudes to the amplitude obtained below 
1.42~\gev\ with our phase shift analyses. 
Although we have verified that the low and high energy amplitudes are compatible, 
within errors, at 1.42~\gev, we have {\sl not} required exact matching. 
The reason for this is that at the lower energy Regge range, say below $\sim1.8\,\gev$, 
some amplitudes still present structure; for example, for the 
amplitudes with isospin unity, the structure associated with 
the $\rho(1450)$, $\rho(1700)$ 
and $\rho_3(1690)$ resonances. 
It is true that these resonances couple weakly to $\pi\pi$, but, at the level of 
precision required in the present paper, this is not negligible:
as happens in the case of $\pi^+ p$ scattering (see e.g. \fig~2 in ref.~11), 
one expects the Regge amplitude to provide only a fit in the mean.
This mismatch produces distortions near the boundary,
 $s^{1/2}=1.42\,\gev$, clearly seen in some of the dispersion relation calculations below;
particularly, for $\pi^0\pi^+$ scattering, where the P and, 
to a lesser extent, the F waves are important.
We have done nothing to correct this distortion which, anyway, only affects 
the points very near 1.42~\gev. 
The alternate possibility, which would be to use phase shift analyses 
up to higher energies, say 1.8~\gev, would only make matters 
worse since it would have to contend with the nonuniqueness and 
unreliability of the experimental data in that 
region, as discussed for example in ref.~15.

\booksection{6. Forward dispersion relations}

\noindent
In this Section we will evaluate forward dispersion relations for the three
independent $\pi\pi$ scattering amplitudes. 
For these calculations we will take the parameters for all partial waves from the fits to
data\fnote{In PY05 we gave two sets of phase shifts and 
inelasticities: one by fitting directly the various sets of experimental data 
(\sect~2 in ref.~1); and a set obtained by requiring, besides fit to data, 
fulfillment of dispersion relations (summarized in Appendix~1 of ref.~1). 
In the present paper we of course {\sl only} use the amplitudes obtained in PY05 by fitting data,
since the ones improved with dispersion relations use a high energy 
($s^{1/2}>0.92\,\gev$) input that is superseded by our 
calculations in the present paper.}
in ref.~1 (PY05), {\sl except} for the S0 and P  waves above 
$0.92\,\gev$, where we use the expressions found in the present paper (for the S0 wave, with the 
K-matrix fit), 
and for the D2 wave, where we take into account the inelasticity above 1.05~\gev. 
For the D0  wave we use the expressions given in
 the present paper all the way from threshold. 

To measure the fulfillment of the dispersion relations we calculate  
the average chi-squared, $\bar{\chi}^2$. This is defined as the sum of 
the
squares of the real part minus 
the result of the dispersive integral, divided by the 
(correlated) errors squared; this we do at energy intervals of 25~\mev,
and divide by the number of points.  Note however, that this average
$\bar{\chi}^2$ does {\it not} come from a fit to the dispersion relations,
but is simply a measure of how well the forward dispersion 
relations are satisfied by the data fits, which are
{\sl independent} for each wave, and independent of dispersion relations.
When calculating this $\bar{\chi}^2$, we first use the parameters for 
phase shifts and inelasticities in PY05; then, we replace the relevant waves 
by the ones in the present paper; and, finally, we also replace the PY05 
Regge parameters with the  ones in Eq.~(5.5).

\topinsert{
\setbox2=\vbox{\hfuzz1truecm{\psfig{figure=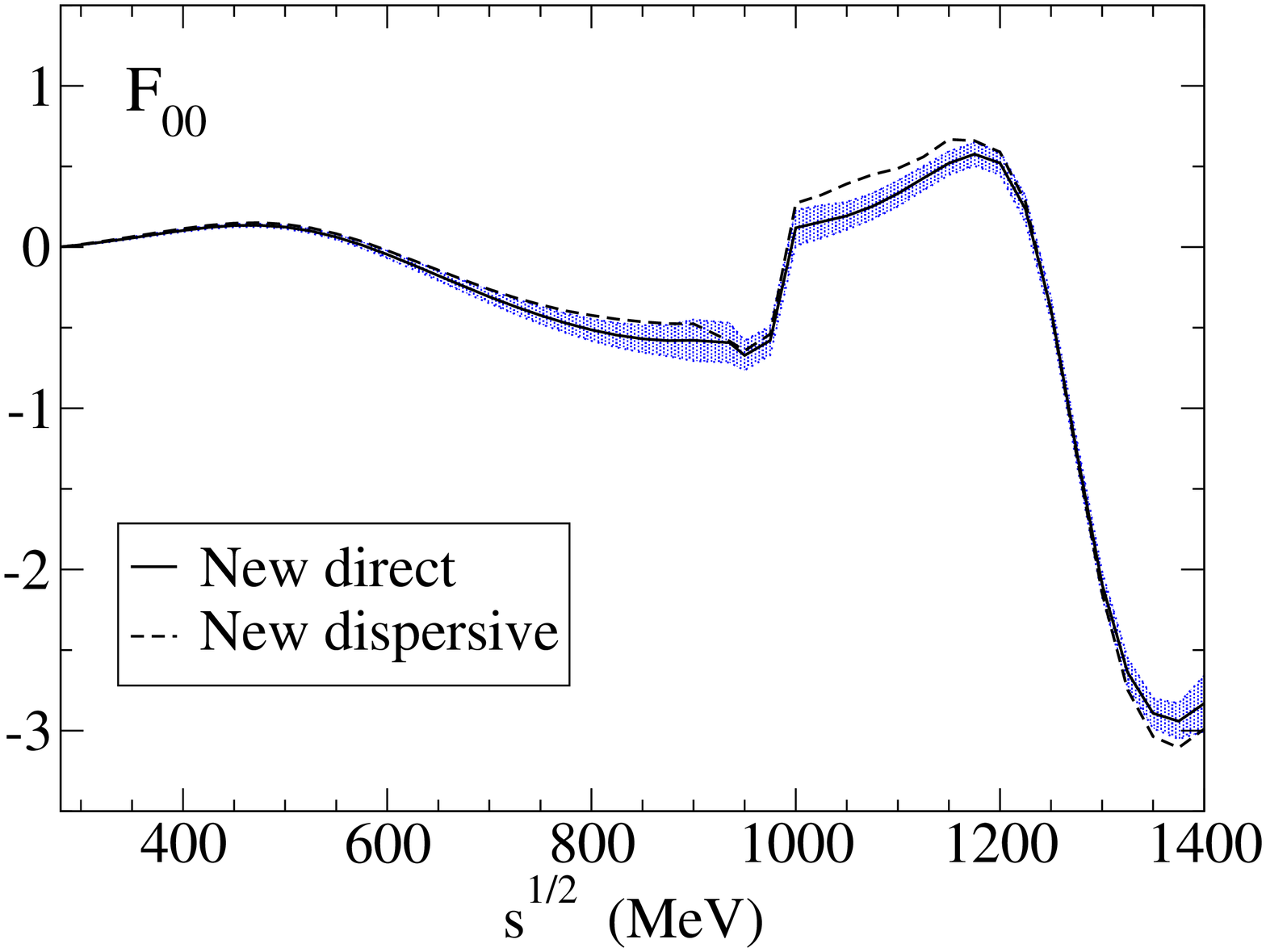,width=9.8truecm,angle=-90}}}  
\setbox3=\vbox{
{\psfig{figure=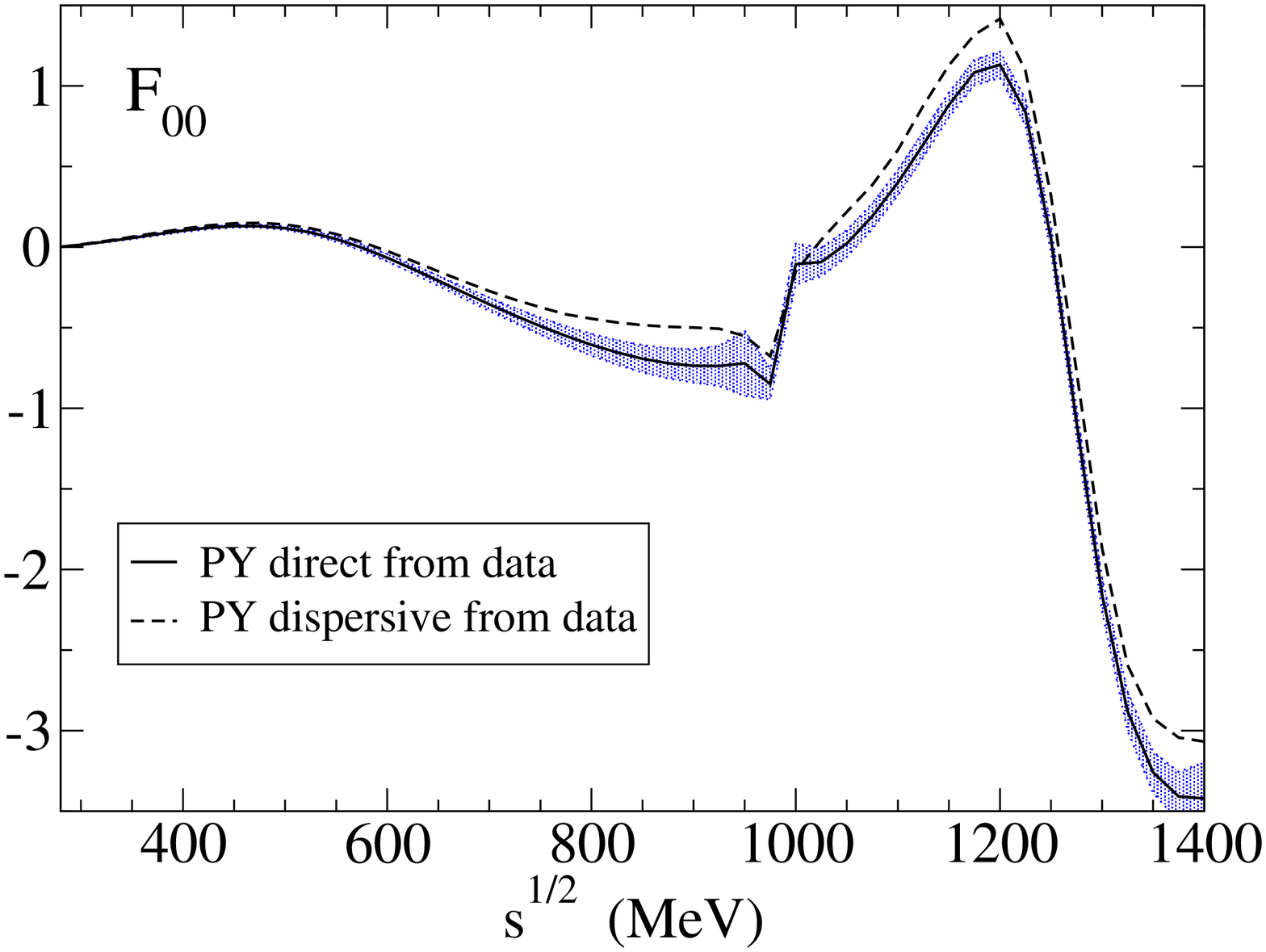,width=9.8truecm,angle=-90}}} 
\setbox6=\vbox{\hsize 5truecm\noindent\petit\figurasc{Figure 6A. }{\hb\hfuzz1.truecm
The $\pi^0\pi^0$ dispersion relation with the new
 S0, D0 and D2 waves.\hb
 Continuous line: real part, evaluated
directly
 with the
parametrizations (the gray band covers the error).\hb
 Dashed line: the result of the dispersive integral, with the 
 Regge parameters of (5.5).\hb\phantom{x}
}}
\setbox7=\vbox{\hsize 5truecm\noindent\petit\figurasc{Figure 6B. }{\hb
The $\pi^0\pi^0$ dispersion relation with the old, PY05
 S0, D0 and D2 waves. Continuous line: real part, evaluated directly
 with the
parametrizations.\hb
 Dashed line: the result of the dispersive integral, with the 
Regge parameters as in  PY05.\hb\phantom{x}
}}
\line{\otightboxit{\box2}\hfil\box6}
\smallskip
\line{\otightboxit{\box3}\hfil\box7}
}\endinsert

\booksubsection{6.1. The $\pi^0\pi^0$ and $\pi^0\pi^+$ dispersion relations}

\noindent
We first evaluate the forward dispersion relation for 
 $\pi^0\pi^0$ scattering, the one that was worse verified in PY05 and the one 
for which the improvement due to the new parametrizations is more marked.
We write
$$\real F_{00}(s)-F_{00}(4M_{\pi}^2)=
\dfrac{s(s-4M_{\pi}^2)}{\pi}\pepe\int_{4M_{\pi}^2}^\infty\dd s'\,
\dfrac{(2s'-4M^2_\pi)\imag F_{00}(s')}{s'(s'-s)(s'-4M_{\pi}^2)(s'+s-4M_{\pi}^2)}.
\equn{(6.1)}$$
The result of the calculation is shown in \fig~6, where the continuous curve is the 
real part evaluated from the parametrizations, 
and the broken curve  is the result of
the dispersive integral, i.e., the right hand side of 
(6.1).

The fulfillment of this dispersion relation improves substantially 
 what we had in PY05:\fnote{Of course, we here compare with 
the results obtained using  the fits to data, {\sl before} improving 
them by requiring fulfillment of the 
dispersion relations at low energy} the changes in the average chi-squared
are 
$$\eqalign{
{\petit\pi^0\pi^0:}\quad&
\hbox{\petit PY05}\;\quad\hbox{\petit New phase sh.}\quad\hbox{\petit New                  
Regge}\cr
\bar{\chi}^2=&\quad 3.8\;\to\;1.52\qquad\quad\quad\to1.41,\quad\qquad\hbox{for}\;
s^{1/2}\leq930\;\mev,\cr
\bar{\chi}^2=&\quad 4.8\;\to\;1.76\qquad\quad\quad\to 1.63,\quad\qquad\hbox{for}\;
s^{1/2}\leq1420\;\mev.\cr}
\equn{(6.2)}$$
Here and in similar expressions below, ``New phase sh." means 
that we use the new, improved phase shifts (and
inelasticities) of the present paper; 
``New Regge" means that we also use the new Regge parameters in (5.5).
In both cases we use  the
K-matrix fit  for the S0 wave, Eqs.~(2.4).

\topinsert{
\setbox2=\vbox{\hfuzz1truecm
{\psfig{figure=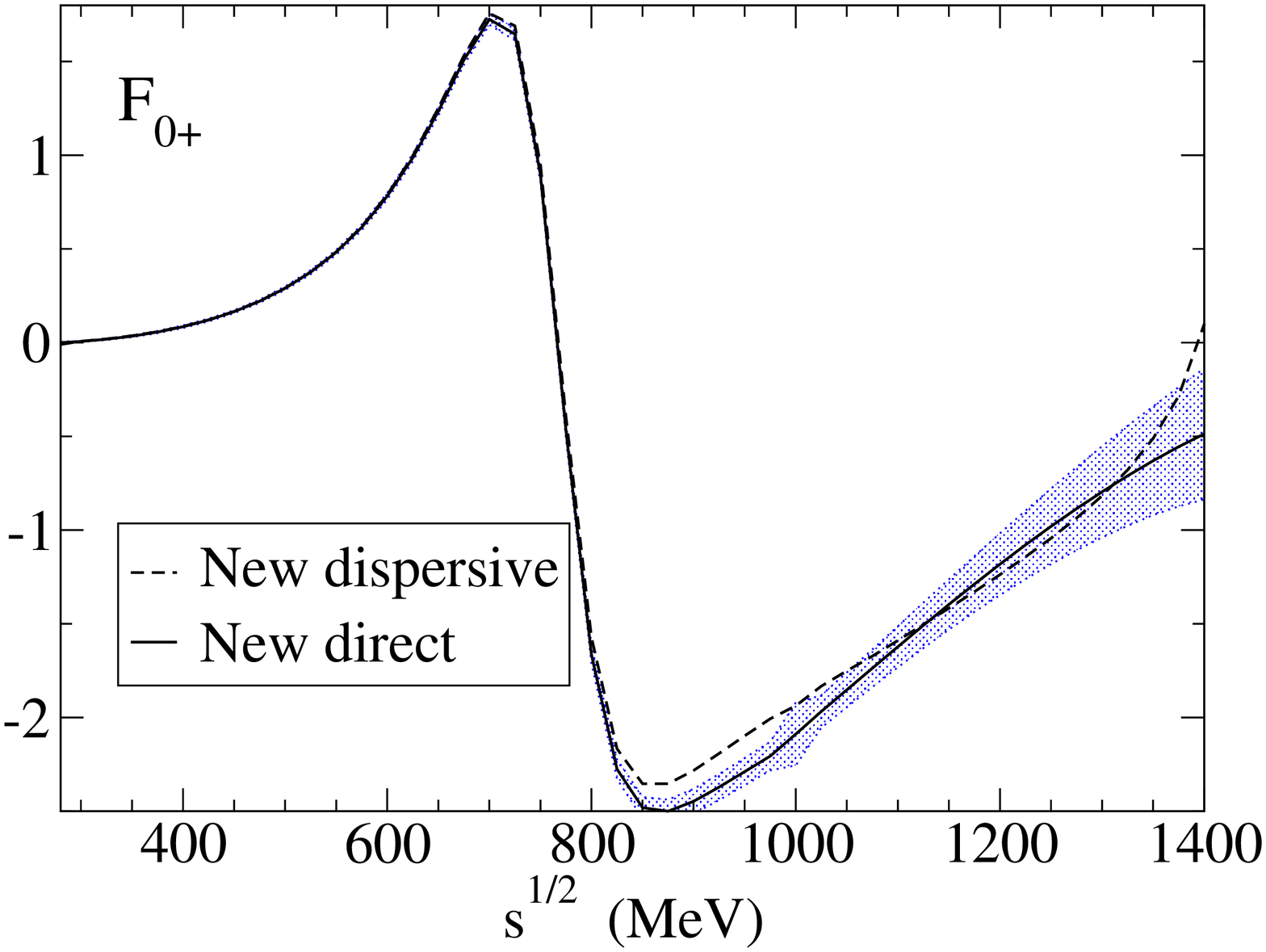,width=9.8truecm,angle=-90}}}  
\setbox3=\vbox{\hfuzz1.truecm
{\psfig{figure=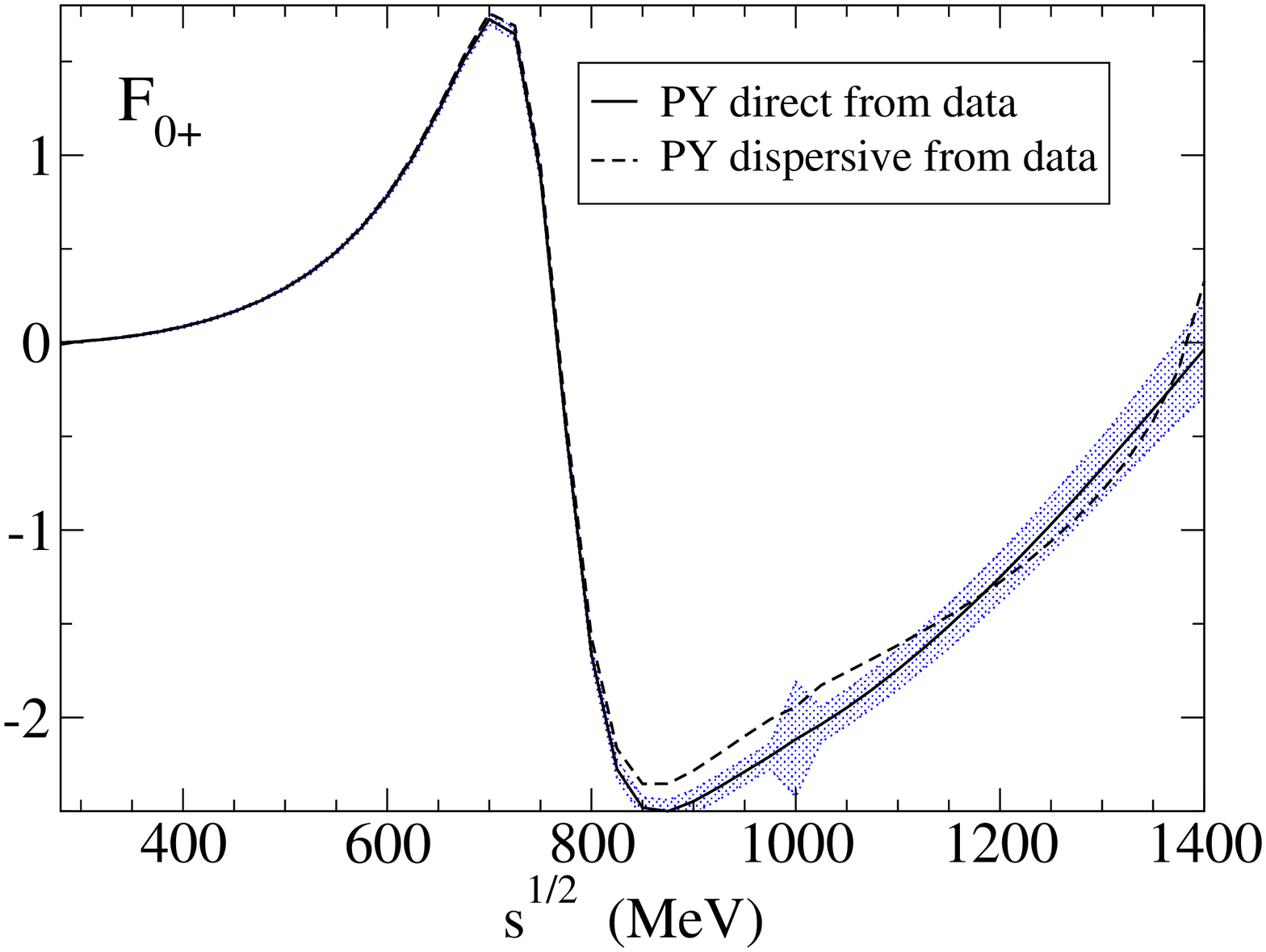,width=9.8truecm,angle=-90}}} 
\setbox6=\vbox{\hsize 5.5truecm\noindent\petit\figurasc{Figure 7A. }{\hb
The $\pi^0\pi^+$ dispersion relation with the
 new P and D2 waves.\hb Continuous line: real part, evaluated 
directly with the
parametrizations.\hb
 Dashed line: the result of the dispersive integral, with the 
 Regge parameters of (5.5).\hb\phantom{x}
}}
\setbox7=\vbox{\hsize 5.5truecm\noindent\petit\figurasc{Figure 7B. }{\hb
The $\pi^0\pi^+$ dispersion relation with the old, PY05
 P and D2  waves.\hb Continuous line: real part, evaluated with the
parametrizations.\hb
 Dashed line: the result of the dispersive integral, with the old, 
PY05 Regge parameters.\hb\phantom{x}
}}
\line{\otightboxit{\box2}\hfil\box6}
\smallskip
\line{\otightboxit{\box3}\hfil\box7}
}\endinsert

The improvement obtained for $\pi^0\pi^0$ when using the 
new phase shifts is more impressive if we remember that the errors 
we have now for the S0 wave above 
0.92~\gev, and for the D0 wave in the whole range, are
substantially  smaller than what we had in PY05.
It is also noteworthy that the improvement in the 
dispersion relation is due almost exclusively to the use of the new
 phase shifts and inelasticities in the range $\sim1$ to 1.42~\gev; 
the improvement due
to introducing the  Regge behaviour (5.5) is much more modest.

The dispersion relation  for  $\pi^0\pi^+$ 
scattering reads, with $F_{0+}(s)$ the forward $\pi^0\pi^+$ amplitude,
$$\real F_{0+}(s)-F_{0+}(4M_{\pi}^2)=
\dfrac{s(s-4M^2_\pi)}{\pi}\pepe\int_{4M_{\pi}^2}^\infty\dd s'\,
\dfrac{(2s'-4M^2_\pi)\imag F_{0+}(s')}{s'(s'-s)(s'-4M_{\pi}^2)(s'+s-4M_{\pi}^2)}.
\equn{(6.3)}$$
In \fig~7 we show the fulfillment of (6.3), both with what we had in PY05 
and with the new phase shifts and 
Regge parameters.

The forward dispersion relation for  $\pi^0\pi^+$ 
scattering was already very well satisfied with 
the parameters in PY05; it becomes slightly 
better satisfied now. 
The changes in the average chi-squared are
$$\eqalign{
{\petit\pi^0\pi^+:}\quad&\hbox{\petit PY05}\;\quad\hbox{\petit New phase sh.}
\quad\hbox{\petit New Regge}\cr
\bar{\chi}^2=&\quad 1.7\;\to\;1.75\qquad\quad\quad\to1.60,\quad\qquad\hbox{for}\;
s^{1/2}\leq930\;\mev,\cr
\bar{\chi}^2=&\quad 1.7\;\to\;1.60\qquad\quad\quad\to1.44,\quad\qquad\hbox{for}\;
s^{1/2}\leq1420\;\mev;\cr}
\equn{(6.4)}$$

The improvement here, although existing, is rather small: 
not surprisingly as the corresponding amplitude does not contain the S0 or D0 waves.
The amelioration is due only to use of the new Regge parameters from Eq.~(5.5).

The fact that both the dispersion relations for $\pi^0\pi^0$ and
 $\pi^0\pi^+$  improve with the present parameters for the $P'$ trajectory 
confirms the correctness of the procedure for 
determining it which we developed in \sect~5.

\topinsert{
\setbox2=\vbox{\hfuzz1truecm{
{\psfig{figure=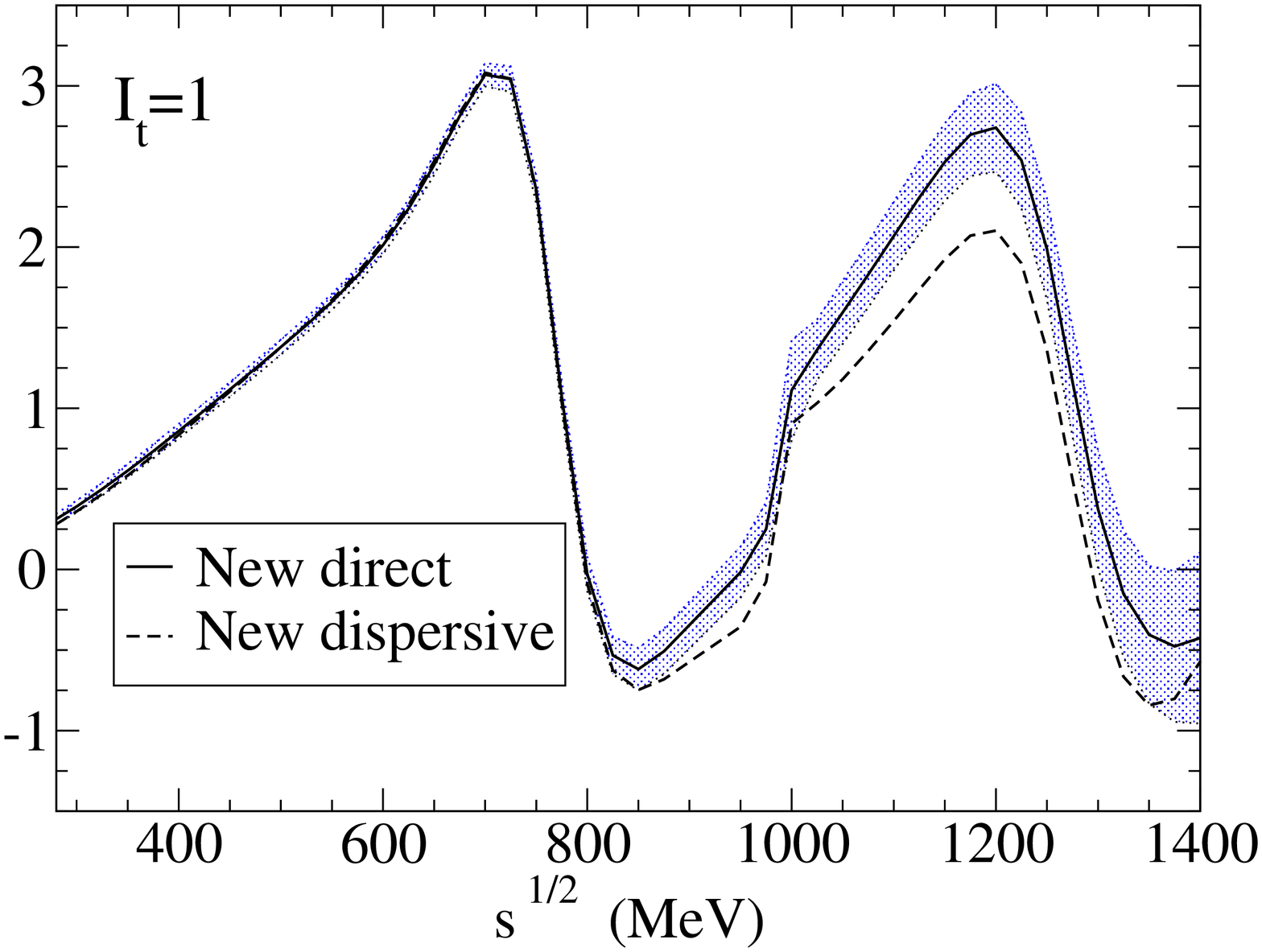,width=9.8truecm,angle=-90}}}}  
\setbox3=\vbox{\hfuzz1truecm
{\psfig{figure=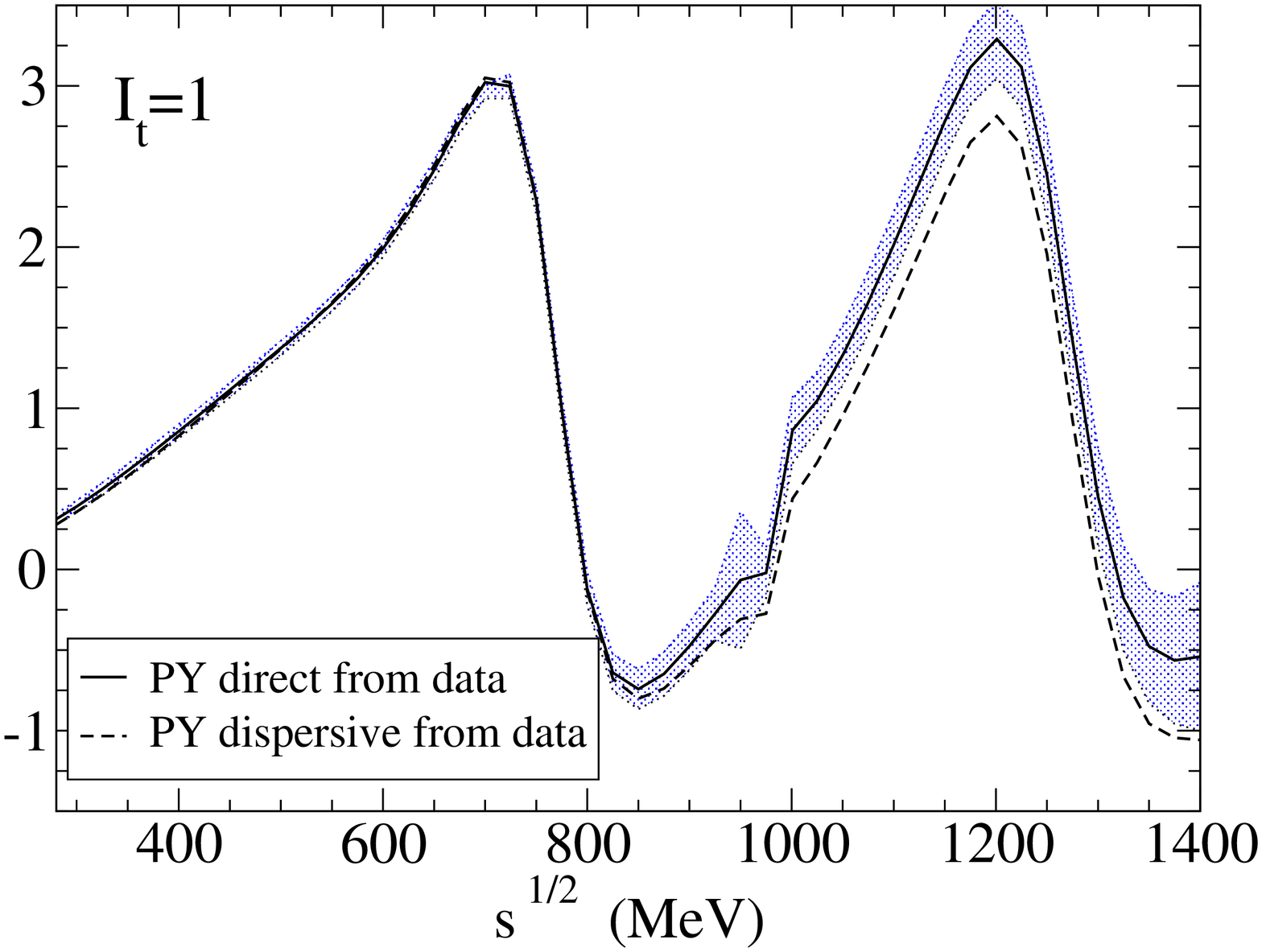,width=9.8truecm,angle=-90}
}} 
\setbox6=\vbox{\hsize 5.5truecm\noindent\petit\hfuzz0.5truecm\figurasc{Figure 8A. }{
The  dispersion relation for the $I_t=1$ amplitude, 
calculated with the new amplitudes.
 \hb
Continuous line: real part and error (shaded area)
 evaluated directly with the parametrizations.\hb
 Dashed line: the result of the dispersive integral, with the  
Regge parameters given in (5.5).\hb\phantom{x}
}}
\setbox7=\vbox{\hsize 5.5truecm\noindent\petit\figurasc{Figure 8B. }{
The  dispersion relation for the $I_t=1$ amplitude with 
the old, PY05 S0, D0 and P waves.\hb
Continuous line: real part and error (shaded area)
 evaluated directly.\hb
 Dashed line: the result of the dispersive integral, 
with the PY05 Regge parameters for the rho.\hb\phantom{x}
}}
\line{\otightboxit{\box2}\hfil\box6}
\smallskip
\line{\otightboxit{\box3}\hfil\box7}
}\endinsert

\booksubsection{6.2. The dispersion relation for the $I_t=1$ scattering 
amplitude}

\noindent
The dispersion relation for the $I_t=1$ scattering 
amplitude does not require subtractions, and  reads 
$$\real F^{(I_t=1)}(s,0)=\dfrac{2s-4M^2_\pi}{\pi}\,\pepe\int_{4M^2_\pi}^\infty\dd s'\,
\dfrac{\imag F^{(I_t=1)}(s',0)}{(s'-s)(s'+s-4M^2_\pi)}. 
\equn{(6.5)}$$
The result of the calculation is shown in \fig~8.

In this case the contribution of the Regge piece is very important, 
although the details only matter in the region above 1~\gev. Here
 the fulfillment 
of the dispersion relation becomes entangled 
with which Regge behaviour one uses; 
particularly since we now have S0 and D0 amplitudes with 
very small errors  above 1~\gev, which is where the detailed 
shape of the 
Regge amplitude has more influence.

 The changes in the $\bar{\chi}^2$
from what we had in PY05 are
$$\eqalign{
{\petit I_t=1:}\quad&
\hbox{\petit PY05}\quad\;\hbox{\petit New phase sh.}\quad\hbox{\petit New Regge}
\cr
\bar{\chi}^2=&\quad
0.2\;\to\;0.57\qquad\quad\quad\to0.32\qquad\qquad\hbox{for}\;
s^{1/2}\leq930\;\mev,\cr
\bar{\chi}^2=&\quad 1.4\;\to\;2.32\qquad\quad\quad
\to1.76\qquad\qquad\hbox{for}\; s^{1/2}\leq1420\;\mev.\cr}
\equn{(6.6)}$$
The conventions are like in (6.2) above.

The dispersion relation deteriorates a little, 
which indicates that the rho Regge parameters may still be improved.
In fact, it is remarkable that the simple change of (5.5) in place of (5.2,~3) improves
so clearly the
  dispersion relation  above 0.9~\gev\ for exchange of isospin unity, 
while leaving it almost unchanged below this energy for all processes.
This confirms that the Regge parameters are much better determined for 
exchange of isospin zero than for exchange of isospin 1, and indicates that a
 complete treatment of dispersion relations (in particular, using them 
to improve  the scattering amplitudes)
 may require simultaneous consideration of the 
Regge parameters and of the  parameters of the phase shift analyses, 
as in fact was done in PY05.  
We will leave this for a forthcoming paper, where we will
also study the improvement of our
parametrizations using the dispersion relations as well as Roy equations.

Finally, we mention here
that, although the 
improvements 
in the present paper only affected the various waves above $\sim1\,\gev$ (with the exception 
of the very small change of the D0 wave below $\bar{K}K$ threshold), there is a  systematic
improvement  of the dispersion relations also below that energy, 
which is a nontrivial test of the consistency of the parametrizations below and 
above $\bar{K}K$ threshold. 

\booksection{7. A brief discussion}

\noindent
The results of the present article show that, if we improve the scattering amplitudes 
above $\sim1$~\gev\ using more reliable data sets that those we had in PY05, the 
ensuing amplitudes verify much better forward dispersion relations, especially 
above $\bar{K}K$ threshold; but also below it.
Forward dispersion relations, particularly for 
 $\pi^0\pi^0$ and
 $\pi^0\pi^+$ scattering, which (as discussed in 
PY05) have important positivity properties, 
constitute a very stringent filter
when used  
to discriminate against spurious parametrizations or calculations, 
as discussed in PY05 and ref.~15. 
The fact that, with the small errors we have now, {\sl all} values for the 
$\bar{\chi}^2$ are below the 1.8 level, implies that a 
small change  
in the parameters would ensure complete fulfillment
(within errors).
However, it is clear that, although small, some alterations 
are to be expected of the various parameters if we require the amplitudes to 
verify dispersion relations at the $\bar{\chi}^2=1$ level, 
which we will do in a forthcoming article.

These changes are forced by the fact 
that the dispersion relations are not 
yet perfectly satisfied. With respect to this, we have three suspects here. 
First of all, we have that the 
experimental data for the D2 wave (which contributes to all 
processes) are of such a kind that our fit 
cannot be very reliable for the phase shift above 
1~\gev, and is almost pure guesswork for the 
inelasticity. In fact, already in PY05 we discovered that 
requiring  fulfillment of the dispersion 
relations, within errors, forces a change by more than 1~$\sigma$ in the 
phase shift parameters for this D2 wave.

The second possible culprit is the inelasticity for the D0 wave. 
Although it fits (by construction) that of the $f_2$ resonance, 
the expression we have used is, probably, too rigid. 
There is unfortunately very little one can do here, 
since the quality of the data does not allow 
an  accurate treatment.

The final possible culprit is the isospin 1 Regge amplitude: there
 is perhaps room for improvement here.
The same is true, albeit to a lesser extent, for the 
amplitudes for $P'$ and for exchange of isospin 2. 
(Alternatively, it may turn out that, once the D2 wave is improved, any change in the Regge 
parameters is unnecessary).

Finally, it is clear that one cannot improve our amplitudes much,
 since  they are quite good to begin with. 
However, and based on the  preliminary results that we have at present, 
we expect to  show,  in a forthcoming article, that 
it is still possible to hone our amplitude analysis
 by requiring fulfillment of the Roy 
equations and, especially, of forward dispersion relations 
over the whole energy range.

\booksection{Appendix A. The K-matrix formalism}

\noindent
The phase shift $\delta_\pi$ and elasticity parameter $\eta$ for the S0  partial wave,
 for $\pi\pi$ scattering,
 are defined as
$$\eqalign{
\hat{f}_{11}(s)=&\sin\delta_\pi\ee^{\ii \delta_\pi},\quad s<4m^2_K;\cr
\hat{f}_{11}(s)=&\dfrac{\eta\,\ee^{2\ii\delta_\pi}-1}{2\ii},\quad s>4m^2_K.\cr
}\equn{(A.1)}$$
We have
changed a little the notation with respect to the main text; thus, $\delta_\pi$ 
is what we called $\delta_0^{(0)}$ before, $\hat{f}_{11}$ 
was called $\hat{f}_0^{(0)}$ in the main text, etc. 
The index (11) in $\hat{f}_{11}$ is a channel index; see below. Also, 
we do not write angular momentum or isospin indices explicitly.
 We  assume here, as in the main text, that there are only two
channels open (which is likely a
 good approximation below $\sim1.25$~\gev, and not too bad up to 1.42~\gev):
$$(11):\quad\pi\pi\to \pi\pi;\qquad (12):\quad\pi\pi\to \bar{K}K;\qquad
(22):\quad\bar{K}K\to \bar{K}K.$$
Because of time reversal invariance, the channels $\pi\pi\to \bar{K}K$ and 
$\bar{K}K\to\pi\pi$ are represented by the same amplitude. 
We then form a matrix, with elements $\hat{f}_{ij}$, $i,\,j=1,2$,
  $\hat{f}_{11}=\hat{f}_{\pi\pi\to\pi\pi}$, etc.:
$${\bf f}=\pmatrix{\hat{f}_{\pi\pi\to\pi\pi}&\hat{f}_{\pi\pi\to\bar{K} K}\cr
 \hat{f}_{\pi\pi\to\bar{K} K}&\hat{f}_{\bar{K} K\to\bar{K} K}}=
\pmatrix{\dfrac{\eta\,\ee^{2\ii\delta_\pi}-1}{2\ii}& 
\tfrac{1}{2}\sqrt{1-\eta^2}\;\ee^{\ii(\delta_\pi+\delta_K)}\cr
\tfrac{1}{2}\sqrt{1-\eta^2}\;\ee^{\ii(\delta_\pi+\delta_K)}&
\dfrac{\eta\,\ee^{2\ii\delta_K}-1}{2\ii}}.
\equn{(A.2)}$$
$\delta_K$ is the phase shift for $\bar{K}K\to \bar{K}K$ scattering.  
Below $\bar{K}K$ threshold, the elasticity parameter is $\eta(s)=1$;
 above  $\bar{K}K$ threshold
 one has the bounds $0\leq\eta\leq1$. 

We write  ${\bf f}$ as
$${\bf f}=\left\{{\bf k}^{-1/2}{\bf K}^{-1}{\bf k}^{-1/2}-\ii\right\}^{-1},
\quad {\bf k}=\pmatrix{k_1&0\cr0&k_2}.
\equn{(A.3)}$$
$k_i$ are the momenta,
$k_1=\sqrt{s/4-M^2_\pi},\quad k_2=\sqrt{s/4-m^2_K}$.
Then, analyticity and unitarity imply that $\bf K$ is analytic 
in $s$ through the 
$\bar{K}K$ threshold; hence, it only depends on $k_2^2$: 
$K_{ij}=K_{ij}(k_2^2)$. 
This is the well-known K-matrix formalism, which the reader may find 
developed in detail in 
the standard textbook of Pilkuhn\ref{16} or, perhaps more accessible, 
in the lecture notes by one of us;\ref{17} and, applied to the S0 wave in 
$\pi\pi$ scattering, in ref.~2.

Because of (A.2,~3), one can express $\delta_\pi$ and $\eta$ in terms of the $K_{ij}$ as
$$
\tan\delta_\pi=\cases{\dfrac{k_1|k_2|\det {\bf K}+k_1K_{11}}{1+|k_2|K_{22}},
\quad s\leq 4m^2_K,\cr
\eqalign{&\,\dfrac{1}{2k_1[K_{11}+k_2^2K_{22}\det {\bf K}]}\Bigg\{
k^2_1K^2_{11}-k^2_2K^2_{22}+k_1^2k_2^2(\det{\bf K})^2-1\cr
+&\,
\sqrt{(k^2_1K^2_{11}+k^2_2K^2_{22}+k_1^2k_2^2(\det{\bf K})^2+1)^2-
4k_1^2k_2^2K^4_{12}}\;\Bigg\},\quad s\geq 4m^2_K;} 
\cr}
\equn{(A.4a)}$$
one also has
$$\eta=\sqrt{\dfrac{(1+k_1k_2\det{\bf K})^2+(k_1K_{11}-k_2K_{22})^2}
{(1-k_1k_2\det{\bf K})^2+(k_1K_{11}+k_2K_{22})^2}},\quad s\geq 4m^2_K.
\equn{(A.4b)}$$
The sign in the surd in (A.4a) is to be taken positive if, as happens in our case,
$K_{11}(k_2^2=0)>0$.

From the relation between the phase shift above and below threshold, and also 
with the elasticity, it may appear that one could write an expansion for 
$\delta_\pi(s)$ below threshold and from it, deduce corresponding expressions for 
$\eta_\pi(s)$ and for $\delta_\pi(s)$ above threshold.
This comes about as follows. Let us define  $\delta^{\rm b}_\pi(s)$ 
to be the phase shift below threshold, and  $\delta^{\rm a}_\pi(s)$  that above threshold, 
both as given in (A.4a). 
Write the Taylor expansion
$$\delta^{\rm b}_\pi(s)=\sum_0^\infty a_n \kappa^n/m_K^n,\quad a_0\equiv d_0
\equn{(A.5)}$$
and $\kappa=|k_2|$ below threshold. 
Substituting this into the expression, valid below threshold,
$$\hat{f}_{\pi\pi\to\pi\pi}=\dfrac{\ee^{2\ii\delta_\pi}-1}{2\ii},\quad (s\leq 4m^2_K)$$
and continuing this across the cut in the variable $\kappa=-\ii k_2$ above the threshold 
we find the expression, valid for $s\geq 4m^2_K$,
$$\hat{f}_{\pi\pi\to\pi\pi}=
\dfrac{\ee^{2(a_1k_2/m_K-a_3k_2^3/m^3_K+\cdots)}
\ee^{2\ii(d_0-a_2k_2^2/m^2_K+a_4k_2^4/m^4_K+\cdots)}-1}{2\ii},\quad
(s\geq 4m^2_K).$$
On comparing with the expression above threshold given in (A.1) we find
$$\eqalign{\delta^{\rm a}_\pi(s)=&\,d_0-a_2k_2^2/m^2_K+a_4k_2^4/m^4_K+\cdots;\cr
\eta(s)=&\,\ee^{2(a_1k_2/m_K-a_3k_2^3/m^3_K+\cdots)},
\quad (s\geq 4m^2_K).
\cr}
\equn{(A.6)}$$
However, the convergence of (A.6) can only be guaranteed in a disk touching  the 
left hand cut of the K-matrix, a cut due to the left hand cut in 
$\bar{K}K\to\pi\pi$ scattering,\fnote{It is not difficult to
check that,  although $\hat{f}_{11}(s)$ or  $\hat{f}_{22}(s)$ have no left hand
 cut above $s=0$, $\eta(s)$ and  $\delta^{\rm a}(s)$ 
do. 
For e.g. the first, we use (A.1) and find
$$\eta^2=
\dfrac{(2\ii\hat{f}_{11}+1)(2\ii\hat{f}_{22}+1)}
{(2\ii\hat{f}_{11}+1)(2\ii\hat{f}_{22}+1)+4\hat{f}^2_{12}}$$
from which it is obvious that $\eta$ inherits the left hand cut of $\hat{f}_{12}$.} that runs up to
$s=4(m^2_K-M^2_\pi)$: therefore, only for
$|k_2|<M_\pi$.  From a practical point of view, we have checked numerically 
 that fitting  with (A.5), (A.6) 
 represent reasonably well $\delta^{\rm b}_\pi(s)$ and 
 $\delta^{\rm a}_\pi(s)$ (this one with irrealistic errors) 
but does certainly not represent
$\eta(s)$,  in the region away from $s=4m^2_K$, unless one adds an inordinately large number of 
parameters.

On the other hand, it is clear  that all three $\delta^{\rm b}_\pi(s)$ ,
 $\delta^{\rm a}_\pi(s)$ and $\eta(s)$ are continuous functions of, 
respectively, $\kappa$, $k_2^2$ and $k_2$. 
Therefore they can be approximated by polynomials in these variables over the whole range,
 even if they are not one the
continuation of the other.

\topinsert{
\setbox0=\vbox{{\psfig{figure=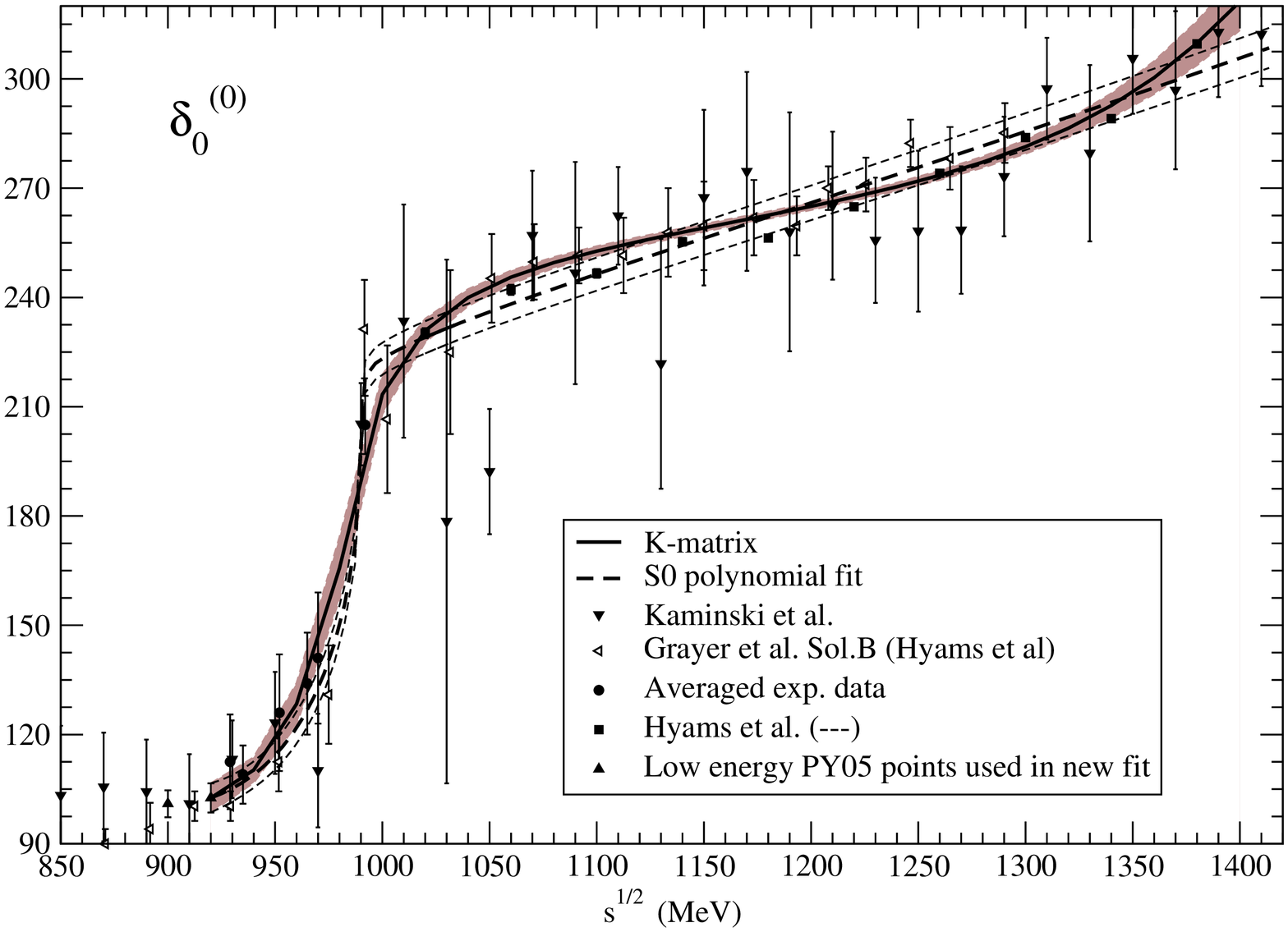,width=14truecm,angle=-90}}} 
\setbox6=\vbox{\hsize 15truecm\captiontype\figurasc{Figure 9A. }{
Comparison of the fits to $\delta_0^{(0)}$: polynomial [Eq.~(B.2)] given by the 
dashed lines, 
and with the K-matrix [Eq.~(2.4)] (solid line and dark area).}\hb} 
\centerline{\otightboxit{\box0}}
\bigskip
\centerline{\box6}
\medskip
}\endinsert

\booksection{Appendix B. Polynomial fit}

\noindent
We present here a polynomial fit to phase shift and elasticity parameter in which 
the three quantities: phase shift below $\bar{K}K$ threshold, phase shift above 
this threshold, and elasticity are fitted separately. 
Although this fit is less reliable than the K-matrix one, especially near
 $\bar{K}K$ threshold, it will allow us to test the importance of multibody channels, not taken
into account in the K-matrix fit.

  For the phase we    write 
$$\delta_0^{(0)}(s)=
\cases{
d_0+a\dfrac{|k_2|}{m_K}+b\dfrac{|k_2|^2}{m^2_K},\quad
(0.92\;\gev)^2<s<4m^2_K;\cr
d_0+B\dfrac{k_2^2}{m^2_K},\quad  4m^2_K<s<(1.42\;\gev)^2;  \quad
k_2=\sqrt{s/4-m^2_K}. \cr}
\equn{(B.1)}$$
The parameters $d_0$, $a$ and $b$ 
are strongly correlated. 
One can get  parameters with low correlation by eliminating the parameter $b$
 in favour of the phase shift
 $d_1$  
at a low energy point, that we 
conveniently take $s^{1/2}=0.92\,\gev$: note that, unlike for the K-matrix fit, 
we now match low and intermediate energy fits at 0.920~\gev. We thus rewrite
the parametrization as
$$\delta_0^{(0)}(s)=
\cases{d_0+a\dfrac{|k_2|}{m_K}+\dfrac{|k_2|^2}{|k_2(0.92^2 \gev^2)|^2}
\left\{d_1-d_0-a\dfrac{|k_2(0.92^2 \gev^2)|}{m_K}\right\},\quad
(0.92\;\gev)^2<s<4m^2_K;\cr d_0+B\dfrac{k_2^2}{m^2_K},
\quad 4m^2_K<s<(1.42\;\gev)^2; 
\quad k_2=\sqrt{s/4-m^2_K}.
\cr}
\equn{(B.2a)}$$
In the previous Appendix~A we presented a discussion about
 these expansions.
 From it it follows that, while the expansion below threshold 
can be considered as convergent in the 
range of interest here, 
$0.92\,\gev\leq s^{1/2}\leq2m_K$, the expansion above threshold
(both for $\delta_0^{(0)}$ and $\eta_0^{(0)}$, see below) 
should be taken as purely phenomenological. In particular, we do {\sl not}
 impose the equality $b=-B$ that would follow if we took (B.2a) to be a Taylor expansion 
(see Eq.~(A.6) in the Appendix).
It is possible to fit requiring $b=-B$, at the cost of adding an extra parameter in 
(B.1), $c|k_2|^3/m^3_K$. The resulting fit is not satisfactory: 
it presents excessively small errors for $s>4m^2_K$, due to the 
forced relation  $b=-B$, which should only be effective near threshold, 
the only region where the expansion converges.

We fit separately data  above and below $\bar{K}K$ 
threshold. The fit returns    
a $\chidof=0.4$ below threshold, and  $\chi^2/{\rm d.o.f.}=0.9$
above threshold;  the values of the parameters are
$$d_0=218.3\pm4.5\degrees,\quad a=-537\pm41\degrees,\quad
d_1=\delta_0^{(0)}(0.920^2\;\gev^2)=102.6\pm4\degrees
\equn{(B.2b)}$$
and
$$ B=96\pm3\degrees. 
\equn{(B.2c)}$$
The resulting phase shift is shown in \fig~9A, compared with the K-matrix fit.

Above $1.25\,\gev$,   the two channel formalism 
is spoiled by the appearance of new channels, notably $\pi\pi\to4\pi$, so 
one does not have an exact connection between the data on $\pi\pi\to\bar{K}K$ 
and $\eta_0^{(0)}$. 
In fact, the numbers one gets for $\eta_0^{(0)}$ from $\pi\pi\to\pi\pi$,
 and those that follow 
from $\pi\pi\to\bar{K}K$,  assuming only two channels, are 
slightly different; see below.
We may take this into account by using a polynomial fit 
(instead of a K-matrix one, as we did in the main text).

We next make a polynomial fit to the elasticity parameter writing
$$\eta_0^{(0)}=1-\left(\epsilon_1\dfrac{k_2}{s^{1/2}}+\epsilon_2\dfrac{k_2^2}{s}+
\epsilon_3\dfrac{k_2^3}{s^{3/2}}\right).
\equn{(B.3a)}$$
In principle, the values of the $\epsilon_i$ are related to the $a,\,c,$\tdots\ 
of (B.1); see Appendix~A, \equn{(A.6)}. 
However, we will {\sl not} impose such relations, but will consider the $\epsilon_i$ as
phenomenological parameters, completely free. 
The reason is that the expansion 
$2(ak_2/m_k+ck_2^3/m^3_K+\cdots)$ is very poorly convergent above $\sim1.2\,\gev$: 
something that is a disaster for $\eta_0^{(0)}$, since the expansion appears 
in an {\sl exponent} (the reason for this divergence, that can be traced to the left hand cut in
$\pi\pi\to\bar{K}K$ scattering, may be found in  Appendix~A). 
 Therefore, we would need to add
extra phenomenological terms,  very large, to compensate for that: 
it is more reasonable to make the fit phenomenological  from the beginning.
What we lose by so doing is that we are overestimating the value of
$\eta_0^{(0)}(s)$ for $s$ very near threshold, 
say for $2m_K<s^{1/2}\lsim0.997\,\gev$,
  a reasonable price to pay to get a good description of the elasticity  
in the rest of the range.

\topinsert{
\setbox0=\vbox{{\psfig{figure=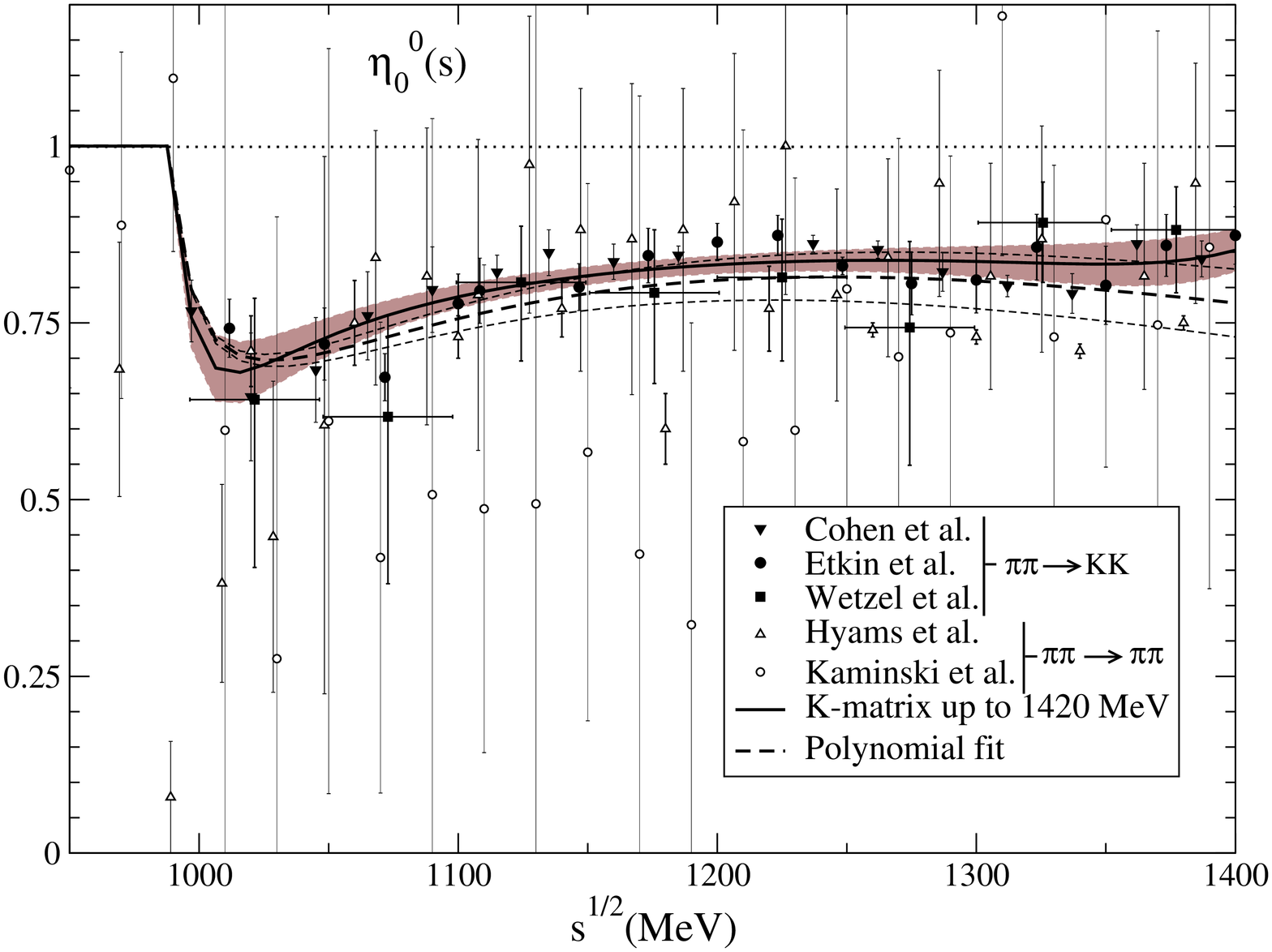,width=12.truecm,angle=-90}}} 
\setbox6=\vbox{\hsize 15truecm\captiontype\figurasc{Figure 9B. }
{Comparison of the fits to $\eta_0^{(0)}$: polynomial [Eq.~(B.3)] given by the dotted lines, 
and with the K-matrix [Eq.~(2.4)] (solid line and dark area).}\hb
\phantom{x}\hb
} 
\centerline{\otightboxit{\box0}}
\medskip
\centerline{\box6}
\medskip
}\endinsert 

We will  consider  the following  possibilities:
 (a) To fit only  $\pi\pi\to\pi\pi$ data 
above 1.25~\gev, in principle the more reliable option;  (b) To fit {\sl also} 
$\pi\pi\to\bar{K}K$
 data above 1.25~\gev; (c) To fit only $\pi\pi\to\bar{K}K$ 
data, in the whole range.
Of course, below 1.25~\gev\ we include both $\pi\pi\to\bar{K}K$
and $\pi\pi\to\pi\pi$ data in the fits (a) and (b).  
We find
$$\eqalign{
\epsilon_1=&\,5.45\pm0.04,\quad\epsilon_2=-30.0\pm0.15,\quad\epsilon_3=46.3\pm0.5;
\quad\hbox{\chidof}=1.1
\qquad\hbox{(a)};\cr
\epsilon_1=&\,5.27\pm0.04,\quad\epsilon_2=-28.2\pm0.15,\quad\epsilon_3=42.2\pm0.5;
\quad\hbox{\chidof}=1.1
\qquad\hbox{(b)};\cr
\epsilon_1=&\,5.77\pm0.05,\quad\epsilon_2=-32.9\pm0.2,\quad\epsilon_3=51.1\pm0.5;
\quad\hbox{\chidof}=0.2
\qquad\hbox{(c)}.\cr}
$$
Note that the errors given here are purely nominal, as the parameters are 
very strongly correlated, while  they were here treated as uncorrelated. 
Note also that the three fits are less separated that it would seem, 
 precisely because of that correlation.
Finally, we remark that the value of  $\eta_0^{(0)}$ that 
follows from 
(c) is {\sl larger} than what follows from (a) or (b); 
and  (b) also slightly above (a). 
These two features constitute very nice consistency tests, since  
taking   $\eta_0^{(0)}$ to be given from  $\pi\pi\to\bar{K}K$ 
as if only two channels were present must surely {\sl underestimate} 
the inelasticity; particularly above $\sim1.2\,\gev$, where 
the process $\pi\pi\to4\pi$ is expected to become nonegligible. 
 
We have verified that one may cover the two fits (a) and (b)
(and even overlap (c), at the edge of the error region) 
by taking as central value that of the fit (a) above and
slightly enlarging the errors. We then get our best result:
$$\epsilon_1=5.45\pm0.06,\quad\epsilon_2=-30.0\pm0.2,\quad\epsilon_3=46.3\pm0.8;
\equn{(B.3b)}$$
 the errors may now be taken as {\sl uncorrelated}.
The resulting elasticity may be seen in Fig.~9B, compared with what we found with the 
K-matrix fit. 
The fact that both determinations overlap 
is a good test of the correctness of our assumption, for the K-matrix fit, 
that the contribution of multiparticle channels is comparable to the error of the fit itself.

The fulfillment of dispersion relations 
with these polynomial fits is just as good as with the K-matrix fit; 
however, the errors of the K-matrix fit are smaller than what we find with the polynomial fit:
 the fulfillment of said dispersion relations may 
therefore be considered to be 
marginally better with the K-matrix formalism, 
which is why we only gave results for the 
dispersion relations  with the K-matrix fit.

\vfill\eject
\booksection{Acknowledgments}

\noindent We are grateful to Profs. I. Caprini and  H. Leutwyler, who communicated us the existence
of an  anomaly in the  K-matrix fit to the S0 wave in a previous version of this work.

 FJY's work
was upported in part by the Spanish DGI of the MEC under contract FPA2003-04597. J.R.P. research is
partially funded by Spanish CICYT contracts FPA2005-02327, BFM2003-00856 as well as Banco
Santander/Complutense contract PR27/05-13955-BSCH, and is part of the EU integrated
infrastructure initiative HADRONPHYSICS PROJECT,
under contract RII3-CT-2004-506078. Finally, 
R.~Kaminski thanks the Complutense University of Madrid for a grant
under the Foreign Doctor Vist Program and the Department of Theoretical Physics II,
where most of his research was carried out.

\booksection{References}

\item{1 }{Pel\'aez, J. R., and Yndur\'ain,~F.~J., {\sl Phys. Rev.} {\bf D71}, 074016
(2005).}
\item{2 }{Hyams, B., et al., {\sl Nucl. Phys.} {\bf B64}, 134, (1973)
 See also the analysis of the 
same experimental data in
Estabrooks, P., and Martin, A. D., {\sl Nucl. Physics}, {\bf B79}, 301, 
(1974).}
\item{3 }{Grayer, G., et al.,  {\sl Nucl. Phys.}  {\bf
B75}, 189, (1974).}
\item{4 }{Protopopescu, S. D., et al., {\sl Phys Rev.} {\bf D7}, 1279, (1973).}
\item{5 }{Kami\'nski, R., Lesniak, L, and Rybicki, K.,
 {\sl Z. Phys.} {\bf C74}, 79 (1997) and 
{\sl Eur. Phys. J. direct} {\bf C4}, 4 (2002).} 
\item{6 }{ Hyams, B., et al., {\sl Nucl. Phys.} {\bf B100}, 205, (1975).}
\item{7 }{ $\pi\pi\to\bar{K}K$ scattering: 
Wetzel,~W., et al., {\sl Nucl. Phys.} {\bf B115}, 208 (1976);
Cohen, ~D. et al., {\sl Phys. Rev.} {\bf D22}, 2595 (1980); 
Etkin,~E. et al.,  {\sl Phys. Rev.} {\bf D25}, 1786 (1982).}
\item{8 }{PDT: Eidelman,  S., et al., {\sl Phys. Letters} {\bf B592}, 1 
 (2004).}
\item{9 }{Wu, F. Q., et al., {\sl Nucl. Phys.} {\bf A735}, 111 (2004).}
\item{10}{de Troc\'oniz, J. F., and Yndur\'ain, F. J., {\sl Phys. Rev.},  {\bf D65},
093001,
 (2002)  
and {\sl Phys. Rev.} {\bf D71}, 073008 (2005).}
\item{11}{Pel\'aez, J.~R., and Yndur\'ain, F. J., {\sl Phys. Rev.} {\bf D69}, 114001 (2004).}
\item{12}{Adel,~K., Barreiro,~F., and Yndur\'ain, F.~J., {\sl Nucl. Phys.} 
{\bf B495}, 221 (1997).}
\item{13}{Rarita, W., et al., {\sl Phys. Rev.} {\bf 165}, 1615, (1968).}
\item{14}{Cudell, J. R., et al., {\sl Phys. Letters} 
{\bf B587}, 78 (2004); Pel\'aez, J.~R., in {\sl Proc. Blois. Conf.
on  Elastic and Diffractive Scattering} (hep-ph/0510005). Note, however, that 
the first reference  fits data for $\pi N$ and $NN$, 
but not for $\pi\pi$, and only for energies above $\sim4\,\gev$; while the last 
article contains only {\sl preliminary} results and, indeed, the parameters for exchange of
isospin zero are not well determined.}
\item{15}{Pel\'aez, J. R., and Yndur\'ain, F. J., {\sl Phys. Rev.} {\bf D68}, 074005 (2003);
 Pel\'aez, J. R., and Yndur\'ain,~F.~J.,
hep-ph/0412320  (Published on the Proc. of the Meeting ``Quark Confinement and 
the Hadron Spectrum", 
Villasimius, Sardinia, September 2004);  Pel\'aez, J. R., and Yndur\'ain, F. J., 
hep-ph/0510216, to appear in the Proc. of the 2005 Montpelier conference.}
\item{16}{Pilkuhn, H.  {\sl The Interaction of 
Hadrons}, North-Holland, Amsterdam, (1967).} 
\item{17}{Yndur\'ain,~F.~J., {\sl Low energy pion physics}, hep-ph/0212282. 
See also Yndur\'ain,~F.~J., {\sl Phys.
Letters} {\bf B612}, 245 (2005).}

\bye